\documentclass[11pt]{article}
\pdfoutput=1
\usepackage{jcapmod}
\usepackage{booktabs}
\usepackage[english]{babel}
\usepackage{amsmath, amssymb, amsfonts, amsbsy, amstext, amsthm}
\usepackage{graphicx}
\usepackage{exscale}
\usepackage[makeroom]{cancel}
\usepackage{soul}
\usepackage{siunitx}
\usepackage{mathtools}
\usepackage{multirow}

\allowdisplaybreaks[1]

\numberwithin{equation}{section}

\def\beq{\begin{equation}}
\def\eeq{\end{equation}}

\def\d{{\rm d}}
\def\H{{\cal H}}
\def\k{{\boldsymbol{k}}}
\def\q{{\boldsymbol{q}}}
\def\x{{\boldsymbol{x}}}
\def\n{{\boldsymbol{n}}}
\def\Nf{N_{\rm eff}}
\def\Nn{N_{\rm fluid}}
\def\c{c_s}
\def\in{{\rm in}}
\def\rec{{\rm rec}}

\DeclareRobustCommand{\SkipTocEntry}[4]{}

\begin{document}

\pagenumbering{roman}
\begin{titlepage}
\baselineskip=15.5pt \thispagestyle{empty}

\bigskip\

\vspace{1cm}
\begin{center}

{\fontsize{20.74}{24}\selectfont \sffamily \bfseries Phases of New Physics in the CMB}

\end{center}
\vspace{0.2cm}
\begin{center}
{\fontsize{12}{30}\selectfont Daniel Baumann,$^{\bigstar}$ Daniel Green,$^{\clubsuit, \blacklozenge}$ Joel Meyers,$^{\clubsuit}$ and Benjamin Wallisch$^{\bigstar}$}
\end{center}

\begin{center}
\vskip 8pt
\textsl{$^\bigstar$ DAMTP, Cambridge University, Cambridge, CB3 0WA, UK}
\vskip 7pt

\textsl{$^\clubsuit$ Canadian Institute for Theoretical Astrophysics, Toronto, ON M5S 3H8, Canada}
\vskip 7pt

\textsl{$^\blacklozenge$ Canadian Institute for Advanced Research, Toronto, ON M5G 1Z8, Canada}
\end{center}

\vspace{1.2cm}
\hrule \vspace{0.3cm}
\noindent {\sffamily \bfseries Abstract} \\[0.1cm]
Fluctuations in the cosmic neutrino background are known to produce a phase shift in the acoustic peaks of the cosmic microwave background.  It is through the sensitivity to this effect that the recent CMB data has provided a robust detection of free-streaming neutrinos.  In this paper, we revisit the phase shift of the CMB anisotropy spectrum as a probe of new physics. The phase shift is particularly interesting because its physical origin is strongly constrained by the analytic properties of the Green's function of the gravitational potential.  For adiabatic fluctuations, a phase shift requires modes that propagate faster than the speed of fluctuations in the photon-baryon plasma. This possibility is realized by free-streaming relativistic particles, such as neutrinos or other forms of dark radiation. Alternatively, a phase shift can arise from isocurvature fluctuations.  We present simple models to illustrate each of these effects. We then provide observational constraints from the Planck temperature and polarization data on additional forms of radiation.  We also forecast the capabilities of future CMB Stage IV experiments. Whenever possible, we give analytic interpretations of our results.
\vskip 10pt
\hrule
\vskip 10pt

\vspace{0.6cm}
\end{titlepage}

\thispagestyle{empty}
\setcounter{page}{2}
\tableofcontents

\clearpage
\pagenumbering{arabic}
\setcounter{page}{1}

%%%%%%%%%%%%%%%%
\section{Introduction}
\label{sec:intro}
%%%%%%%%%%%%%%%%

Cosmology is a sensitive probe of physics beyond the Standard Model (BSM).  The temperature in the early universe was high enough to make the production of weakly interacting and/or massive particles efficient.  If the energy density carried by these particles was significant, then even their gravitational influence can be detected.  This sensitivity to extremely weakly interacting particles is a unique advantage of cosmological probes of BSM physics.  

\vskip 4pt
Until recently, the strongest cosmological constraints on BSM physics came from Big Bang nucleosynthesis~(BBN).  Measurements of the primordial abundances place stringent constraints  both on additional forms of radiation and on injections of energy around the time of BBN (see e.g.~\cite{Sarkar:1995dd, Iocco:2008va, Pospelov:2010hj, Cyburt:2015mya} for reviews).  These constraints are now complemented by the precise observations of the cosmic microwave background (CMB).  Measurements of the temperature anisotropies have been used to test a wide variety of models (most recently by the Planck collaboration~\cite{Ade:2015xua}) ranging from additional sources of radiation, to dark matter annihilation~\cite{Padmanabhan:2005es} and neutrino masses~\cite{Dolgov:2002wy, Lesgourgues:2006nd}.  In the future, CMB polarization experiments will play an increasingly important role in the search for BSM physics. Forecasts for a CMB Stage~IV (CMB-S4) experiment~\cite{Abazajian:2013oma} suggest an increase in sensitivity by one or two orders of magnitude within the next decade.  With such remarkable improvements in sensitivity on the horizon, we should re-assess how this data could sharpen our understanding of the early universe, and, particularly, how it will inform our view of extensions of the standard models of particle physics and cosmology.

\begin{figure}[h!t]
\begin{center}
\includegraphics[scale=0.65]{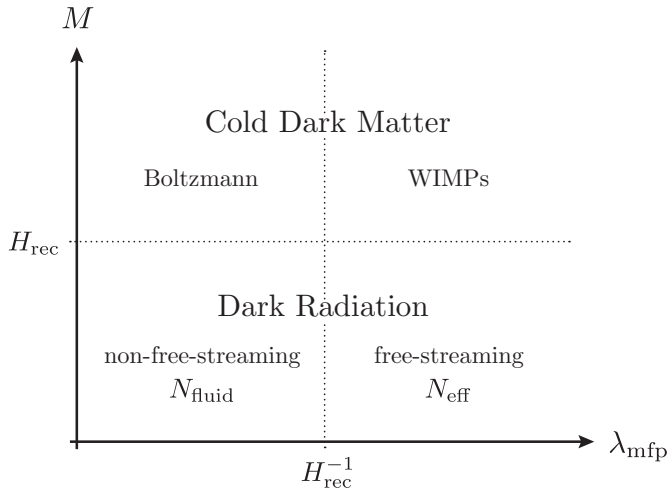}
\caption{Particles beyond the Standard Model can be classified according to their masses $M$ and their mean free paths $\lambda_{\rm mfp}$ (both normalized relative to the Hubble rate at recombination, $H_\rec$). Particles with $M > H_\rec$ contribute to the cold dark matter of the universe, while particles with $M < H_\rec$ are relativistic at recombination. Massive, strongly interacting particles are Boltzmann-suppressed and, therefore, do not contribute a cosmologically interesting density. Dark radiation separates into free-streaming  and non-free-streaming particles. Note that axions, and other non-thermal relics, escape the simple characterization of this figure.}
\label{fig:BSM}
\end{center}
\end{figure}

\vskip 4pt
In this paper, we will revisit the analytic treatment of the CMB anisotropies with an eye towards BSM applications. While numerical codes are ultimately necessary in order to make precise predictions for any particular model, analytic results still play a vital role.  It is through the physical understanding of the data that we can devise new tests and motivate new models.  For example, the use of B-modes in the search for primordial gravitational waves arose from a clear analytic understanding of CMB polarization~\cite{Seljak:1996ti, Seljak:1996gy, Zaldarriaga:1996xe, Kamionkowski:1996ks}.  Similarly, we wish to identify CMB observables that are sensitive to well-motivated forms of BSM physics, but are not strongly degenerate with other cosmological parameters. We will advocate the phase shift\hskip 1pt\footnote{This phase shift refers to a constant shift in the locations of the high-$\ell$ acoustic peaks.  We emphasize that this is a distinct effect from the locations of the first few acoustic peaks which are sensitive to many cosmological parameters, as studied e.g.~in~\cite{Hu:2000ti, Doran:2001yw, Corasaniti:2007rf}.} of the acoustic peaks of the CMB spectrum~\cite{Bashinsky:2003tk} as an observable with the desired characteristics. The physical conditions that lead to a phase shift are rather restrictive and determined by the analytic properties of the Green's function of the gravitational potential. For adiabatic fluctuations, a phase shift requires fluctuations that travel faster than the sound speed of the photon-baryon fluid. This arises naturally for free-streaming relativistic particles, such as neutrinos (but also for other forms of ``dark radiation").  Alternatively, a phase shift can also arise from isocurvature fluctuations.  The phase shift therefore probes an interesting regime in the parameter space of BSM models (see Fig.~\ref{fig:BSM}).

\vskip 4pt
While particle physics experiments give strong constraints on specific scenarios, they can be blind to unknown or incompletely specified forms of new physics.  In contrast, cosmological observations can provide broad constraints on phenomenological parameterizations. This has the advantage of compressing large classes of BSM physics into broad categories and is less sensitive to the details of the models.  (This is analogous to the use of {\it simplified models} to search for new physics at the LHC~\cite{Alves:2011wf}, where one reduces large numbers of models to a single model which captures their essential features.)  This approach has led to important discoveries in the past: by comparing observations against simple phenomenological parameterizations, the existence of dark matter ($\Omega_m$) and dark energy ($\Omega_\Lambda$) was established, the baryon asymmetry ($\eta$) was identified, and evidence for cosmological inflation ($n_s$) was presented.  

\vskip 4pt
A useful way of parameterizing the effects of extra light species on the CMB is in terms of the effective number of free-streaming species, $X$, and non-free-streaming species, $Y$. The former is conventionally defined as the effective number of neutrinos,\footnote{Stated imprecisely, colliders also constrain the number of neutrino species through precision measurements of the width of the $Z$ decay.  Yet, when stated more carefully, collider measurements only tell us how many fermions with mass below $\frac{1}{2}m_Z$ couple to the $Z$ boson \cite{ALEPH:2005ab}.}
\beq
\Nf \equiv  a_\nu \, \frac{\rho_{X}}{\rho_\gamma}\, ,  \label{eq:Neff}
\eeq
where $a_\nu \equiv  \frac{8}{7} \left(\frac{11}{4}\right)^{4/3}$.  Here, $\rho_X$ includes ordinary SM neutrinos (with $N_\nu = 3.046$~\cite{Mangano:2001iu}), but also characterizes any free-streaming radiation density beyond the SM expectation (including additional energy in neutrinos). The current constraint from the Planck satellite, $\Nf =  3.15 \pm 0.23$~\cite{Ade:2015xua}, represents a highly significant detection of the cosmic neutrino background.  Furthermore, these measurements put strong limits on many extensions of the Standard Model containing additional light fields and/or thermal histories that enhance or dilute the energy in neutrinos~\cite{Jungman:1995bz, Cadamuro:2010cz, Menestrina:2011mz, Boehm:2012gr, Brust:2013ova, Weinberg:2013kea, Cyr-Racine:2013fsa, Vogel:2013raa, Millea:2015qra, Chacko:2015noa}.\footnote{For thermally-decoupled massless particles, the equivalent change to $\Nf$ will be a function of the decoupling temperature. A single thermally-decoupled species implies $\delta \Nf \equiv \Nf - 3.046 > 0.02$~\cite{Brust:2013ova}. Interestingly, this level is within reach of future CMB-S4 experiments~\cite{Abazajian:2013oma}.}  In addition, we will allow for a contribution from non-free-streaming radiation.  We capture this by the following parameter:\footnote{Another attempt to parametrize free-streaming radiation is in terms of a viscosity parameter $c_{\rm vis}$~\cite{Hu:1998kj}.  This parameter has recently been detected by Planck, $c_{\rm vis}^2= 0.331\pm 0.037$~\cite{Ade:2015xua} (see also~\cite{Archidiacono:2013lva, Audren:2014lsa}).  However, as discussed in~\cite{Sellentin:2014gaa}, $c_{\rm vis}^2 =\frac{1}{3}$ is not equivalent to free-streaming radiation and differs from $\Lambda$CDM by $\Delta\chi^2= 20$.  Our parameterization has the advantage that it reproduces $\Lambda$CDM when $\Nf = N_\nu = 3.046$ and $\Nn = 0$. A similar parameterization has appeared in~\cite{Bell:2005dr, Chacko:2015noa,Friedland:2007vv}, and was analyzed with WMAP data in~\cite{Bell:2005dr,Friedland:2007vv}.  However, it has only recently become possible to distinguish these parameters with high significance.} 
\beq
\Nn \equiv  a_\nu \, \frac{\rho_{Y}}{\rho_\gamma}\, .	\label{eq:Neff2}
\eeq
We characterize the influence of $\Nf$ and $\Nn$ on the photon-baryon fluid, and study their distinct imprints in the CMB.  We also describe the types of BSM models that are being probed by these parameters.

\vskip 4pt
Until recently, CMB observations were not sensitive enough to distinguish between $\Nf$ and $\Nn$. Both types of species contribute equally to the background density of the universe and, therefore, affect the CMB damping tail in the same way~\cite{Hou:2011ec}. To separate $\Nf$ and $\Nn$ requires measuring subtle differences in the evolution of perturbations. Free-streaming particles (like neutrinos) create significant anisotropic stress which induces a characteristic phase shift in the CMB anisotropies~\cite{Bashinsky:2003tk}. This phase shift has recently been detected for the first time~\cite{Follin:2015hya}. As we will show, non-free-streaming particles (e.g.~\cite{Cyr-Racine:2013jua, Archidiacono:2013dua, Oldengott:2014qra, Buen-Abad:2015ova, Chacko:2015noa}), in general, do not produce a phase shift (at least as long as the fluctuations are adiabatic and their sound speed is not larger than that of the photons).

Guided by our analytic understanding, we will explore the sensitivity to these effects with the Planck satellite and with a future CMB-S4 experiment, focusing on the ability to distinguish the parameters $\Nf$ and $\Nn$.  Our analysis of the Planck temperature and polarization data leads to the following constraints:\footnote{The constraints assume that the helium fraction $Y_p$ is fixed by consistency with BBN. Results that marginalize over $Y_p$ are presented in \textsection\ref{ssec:MCMC}.}
\beq
\Nf = 2.80^{+0.24}_{-0.23}\, \ (1\sigma)\, , \qquad \Nn < 0.67 \, \ (2\sigma) \, .
\eeq
We see that the current data is already sensitive to the free-streaming nature of the fluctuations. We will explain the important role played by the polarization data in breaking the degeneracy between $\Nf$ and $\Nn$, as well as that with the helium fraction $Y_p$.  We will also show that a CMB-S4 experiment would improve these constraints by up to an order of magnitude under a number of experimental configurations. We will highlight how present and future constraints are driven by measurements of the phase shift. We will also explore how these measurements may be optimized, including through the use of delensing to sharpen the acoustic peaks.
  
\vskip 10pt
The outline of the paper is as follows.  In Section~\ref{sec:analytics}, we derive analytically the effects of new relativistic particles on the perturbations of the photon density. We identify the precise physical conditions that produce a phase shift in the CMB anisotropy spectrum.  We illustrate these effects through an exactly solvable toy model. We also compute the phase shift for a simple model with isocurvature fluctuations and for free-streaming relativistic particles. In Section~\ref{sec:analysis}, we confirm some of these pen-and-paper results through a numerical analysis.  We present new constraints on dark radiation from the Planck 2015 data~\cite{Aghanim:2015wva} and forecast the capabilities of future CMB-S4 experiments~\cite{Abazajian:2013oma}.  Section~\ref{sec:conclusions} contains our conclusions and a description of plans for future work.  In two appendices, we comment on the inclusion of matter (Appendix~\ref{app:ext}) and polarization (Appendix~\ref{app:pol}) in our analytic treatment.

\subsubsection*{Notation and Conventions}

We will work in conformal Newtonian gauge
\beq
\d s^2 = a^2(\tau) \left[(-1-2\Phi) \d \tau^2 + (1-2 \Psi) \delta_{ij} \d x^i \d x^j \right] \, ,
\eeq
where $\tau$ is conformal time.  We use $\tau_0$ for the present time, $\tau_{\rm rec}$ for the time of recombination, $\tau_{\rm eq}$ for matter-radiation equality and $\tau_\in$ for the time at which we set the initial conditions.  We will use a subscript `$\alpha$' to denote quantities evaluated at the time $\tau_\alpha$. The conformal Hubble parameter is $\H \equiv \dot a/a$, where overdots stand for derivatives with respect to $\tau$. It will be convenient to define the sum and the difference of the two metric potentials,
\beq
\Phi_\pm \equiv \Phi \pm \Psi \, .
\eeq
We will use $\rho$ and $P$ for density and pressure, and $\sigma$ for the scalar potential of the anisotropic stress.  Individual components (like photons, matter, neutrinos, etc.)~will be denoted by a subscript $a=\gamma,m,\nu, \cdots$, where $X$ and $Y$ will stand for  free-streaming and non-free-streaming species, respectively.  We use overbars for homogeneous background quantities, $\bar \rho_a$, and write perturbations as $\delta \rho_a \equiv \rho_a- \bar \rho_a$.  Following~\cite{Bashinsky:2003tk}, we define the overdensity in the particle number with respect to both the proper volume, $\delta_a\equiv\delta\rho_a/(\bar \rho_a+ \bar P_a)$, and the coordinate volume,~$d_a$. The relation between the two definitions is $d_a \equiv \delta_a - 3 \Psi$.  The equation of state and the speed of sound are $w_a \equiv \bar P_a/\bar\rho_a$ and $c_a^2 \equiv \delta P_a/\delta\rho_a$, respectively.

\clearpage
%%%%%%%%%%%%%%%%%%%%%%%
\section{Physical Origin of the Phase Shift}
\label{sec:analytics}
%%%%%%%%%%%%%%%%%%%%%%%

\begin{figure}[h!t]
\begin{center}
\includegraphics[scale=0.55]{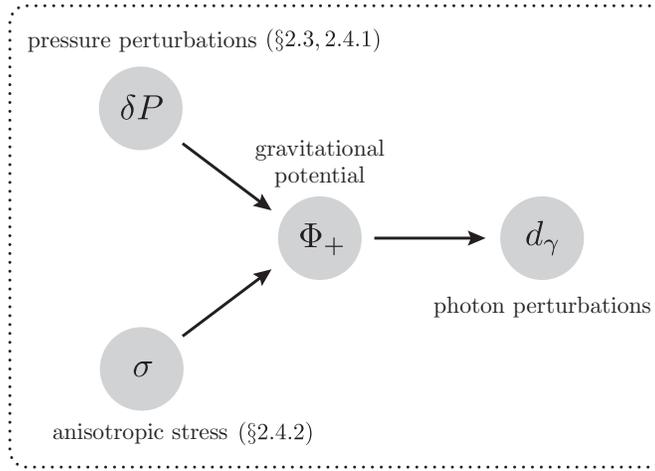}
\caption{Illustration of the coupled perturbations in the primordial plasma.} 
\label{fig:schematic}
\end{center}
\end{figure}

The structure of the acoustic peaks in the CMB is largely determined by the propagation of fluctuations in the photon-baryon plasma.  The physics is that of a harmonic oscillator with a time-dependent gravitational forcing, 
\beq
\ddot d_\gamma - c_\gamma^2\hskip 2pt \nabla^2 d_\gamma = \nabla^2 \Phi_+\, , \label{eq:dgamma}
\eeq
where $c_\gamma^2 \approx \frac{1}{3}$.  A non-trivial evolution of $\Phi_+$ is sourced either by anisotropic stress $\sigma$ or by pressure perturbations~$\delta P$ (see Fig.~\ref{fig:schematic}).  Under certain conditions, which we will identify, this induces a contribution to $d_\gamma$ which is out of phase with its freely oscillating part.  In this section, we will give an analytic description of these effects, building on the pioneering work of Bashinsky\,\&\,Seljak~\cite{Bashinsky:2003tk}.

\subsection{Preliminaries}
\label{sec:EoM}

We begin by collecting a few standard results from cosmological perturbation theory (see e.g.~\cite{Ma:1995ey, Bashinsky:2003tk, Malik:2008im} for further details).  This mainly serves to fix our notation and to introduce the main equations used in this paper.

\vskip 4pt
The CMB couples gravitationally to perturbations in the matter fluctuations.  We define the stress-energy tensor for each species $a$ as
\beq
T^0{}_{0,a} = - (\bar \rho_a+\delta\rho_a)\, , \quad T^0{}_{i,a}  = (\bar \rho_a+ \bar P_a) v_{i, a}\, , \quad T^i{}_{j,a} = (\bar P_a + \delta P_a)\delta^i_j + (\bar \rho_a + \bar P_a) \Sigma^i{}_{j,a} \, .	\label{eq:stressEnergyTensor}
\eeq
The scalar part of the velocity can be written as $v_{i,a} = -\nabla_i u_a$, where $u_a$ is the velocity potential. Similarly, the anisotropic stress tensor $\Sigma_{ij,a}$ can be expressed as $\Sigma_{ij,a} =\frac{3}{2} (\nabla_i\nabla_j - \frac{1}{3} \delta_{ij} \nabla^2) \sigma_a$, where the factor of $\frac{3}{2}$ was introduced for future convenience.  Conservation of the stress-energy tensor for each decoupled species implies
\begin{align}
\dot \delta_a 	&\,=\, \nabla^2 u_a+ 3 \dot \Psi\, , 	\label{eq:C1} \\
\dot u_a 		&\,=\,c_a^2 \delta_a - \chi_a u_a +  \nabla^2 \sigma_a+ \Phi\, ,	\label{eq:C2}
\end{align}
where $\chi_a \equiv \H(1-3 c_a^2)$ is the Hubble drag rate. Equations~\eqref{eq:C1} and \eqref{eq:C2} can be combined into
\beq
\ddot d_a + \chi_a \dot d_a - c_a^2 \nabla^2 d_a = \nabla^4 \sigma_a + \nabla^2(\Phi + 3c_a^2 \Psi)	\, .  \label{eq:perturbationsEoM}
\eeq
To solve this equation requires additional equations for $\sigma_a$ and $\Phi_\pm$.  

\subsubsection*{Boltzmann Equation}

An evolution equation for $\sigma_a$ can be derived from the Boltzmann equation for the distribution functions $f_a(\tau, \x, q, \hat \n)$ of each particle species $a$ with comoving momenta $\q=q\hskip 1pt \hat\n$.  We separate $f_a$ into a background component $\bar f_a$ and a perturbation $\delta f_a \equiv f_a - \bar f_a$. For massless particles, it will be convenient to integrate out the momentum dependence of the distribution function and define 
\beq
\int \d q\, q^3 \hskip 1pt \left(\delta f_a + q \hskip 1pt \partial_q \bar f_a \Psi \right) \ =\ \frac{4}{3}D_a(\tau, \x, \hat \n) \times\! \int \d q\, q^3 \bar f_a(q) \, .
\label{eq:distribution}
\eeq
 The linearized, collisionless Boltzmann equation  is then given by
\beq
\dot D_a+ \hat n^i \nabla_i D_a = -3 \hat n^i \nabla_i\Phi_+ \, .  \label{eq:Boltzmann}	
\eeq
We note that $D_a$ only depends on $\Phi_+ $, but not on $\Phi_- $.  It will be useful to expand the momentum-integrated distribution function $D_a$ into multipole moments,
\beq
D_a = \sum_{\ell=0}^\infty (-i)^\ell (2\ell+1) D_{a,\ell} \hskip 1pt P_{\ell}(\mu)  \, ,	\label{eq:Dmultipoles}
\eeq
where the Legendre polynomials $P_{\ell}(\mu)$ are functions of $\mu=\hat\n \cdot \hat\k$. The monopole moment $D_{a,0}$ determines the overdensity $d_a$, while the quadrupole moment $D_{a,2}$ is associated with the anisotropic stress $\sigma_a$. To see this, one writes the perturbed stress-energy tensor in terms of the perturbed distribution function,
\beq
\delta T^\mu{}_{\nu,a} = a^{-4} \int \d\Omega_{\hat n}\, \hat n^\mu \hat n_\nu \! \int \d q\, q^3 \hskip 1pt \delta f_a \, .
\eeq
Comparing this expression to (\ref{eq:stressEnergyTensor}), we find
\beq
D_{a,0} = d_a \, , \quad \ D_{a,1} = k u_a \, , \quad \ D_{a,2} = \frac{3}{2} k^2 \sigma_a \, .	\label{eq:DmultipoleLinks}
\eeq
The quadrupole moment of~(\ref{eq:Boltzmann}) then provides the missing evolution equation for the anisotropic stress.

\subsubsection*{Einstein Equations}

The metric potentials $\Phi$ and $\Psi$ are determined by the Einstein equations
\begin{align}
\ddot\Psi + \H(2\dot\Psi + \dot\Phi) + (2\dot\H+\H^2)\Phi + \frac{1}{3}\nabla^2(\Phi-\Psi) &= 4 \pi G a^2 \delta P \, , \label{equ:evolution}\\[4pt]
\nabla^2 \Psi - 3\H(\dot\Psi + \H \Phi) &= 4\pi G a^2\, \delta \rho  \, , \label{eq:Poisson}
\end{align}
where $\delta\rho \equiv \sum_a \delta \rho_a$ and $\delta P \equiv \sum_a \delta P_a$ are
the total density and pressure perturbations, respectively. In terms of the fields $\Phi_\pm$, the evolution equation~(\ref{equ:evolution}) becomes
\beq
\ddot \Phi_+ + 3 \H \dot \Phi_+ + (2\dot \H + \H^2) \Phi_+ \,=\, 8\pi G a^2\, \delta P \,+\, {\cal S}[\Phi_-] \, , \label{eq:Einstein0}
\eeq
where ${\cal S}[\Phi_-] \equiv \ddot\Phi_- + \H\dot\Phi_- - (2\dot\H+\H^2 + \frac{2}{3}\nabla^2)\Phi_- $.  The field $\Phi_-$ is related to the total anisotropic stress $\sigma$ via the constraint equation 
\beq
\Phi_- = - 12\pi G a^2\, (\bar \rho+\bar P) \sigma\, , \label{eq:EinsteinC}
\eeq
where $(\bar \rho + \bar P)\hskip 1pt \sigma \equiv \sum_a (\bar \rho_a + \bar P_a)\hskip 1pt \sigma_a$.  In the standard model, both photons and neutrinos contribute to $\delta P$, but only neutrinos provide a source for $\sigma$ (and hence $\Phi_-$).  BSM particles may lead to additional pressure and/or anisotropic stress. 

\vskip 4pt
During radiation domination, the evolution equation~(\ref{eq:Einstein0}) can be written as
\beq
\Phi_+'' + \frac{4}{y} \Phi_+' + \Phi_+ \,=\,  \frac{8\pi G a^2}{(c_\gamma k)^2} \sum_{a} (c_a^2-c_\gamma^2)\, \delta\rho_a \,+\, \tilde {\cal S}[\Phi_-] \, , \label{eq:Phi+RD}
\eeq
where primes denote derivatives with respect to $y \equiv c_\gamma k \tau$ and $\tilde {\cal S}[\Phi_-] \equiv \Phi_-'' + (2/y)\Phi_-' +3\Phi_-$.  The Green's function for~(\ref{eq:Phi+RD})~is
\beq
G_{\Phi_+}(y, y') = \Theta(y-y') \frac{y'}{y^3} \Big[ (y' -y)\cos(y' -y) - (1+ y y') \sin(y' -y) \Big]\, , \label{eq:Gplus}
\eeq
where $\Theta$ is the Heaviside function.  Given a model for the sources in (\ref{eq:Phi+RD}), we use the Green's function to determine the solution for $\Phi_+$.  The time evolution of the source terms may require solving the Boltzmann equation~(\ref{eq:Boltzmann}).

\subsection{Conditions for a Phase Shift}
\label{sec:originPhaseShift}
  
We are now in the position to analyze the evolution of perturbations in the photon-baryon plasma.  For simplicity, we will ignore the small effect due to the baryons,\footnote{We ignore the contributions of baryons and dark matter to the energy density, but we are implicitly including the baryons when we assume that the photons are not free-streaming particles.} so that the speed of fluctuations in the photon density is $c_\gamma^2 \approx \frac{1}{3}$. The Hubble drag rate in~(\ref{eq:perturbationsEoM}) therefore vanishes, $\chi_\gamma \approx 0$, and we get the evolution equation~(\ref{eq:dgamma}).  The solution for $d_\gamma$ can then be written as
\beq
d_\gamma(y) = d_{\gamma,\in} \cos y - c_\gamma^{-2} \int^y_{0}  \d y'\, \Phi_+(y') \sin(y- y')\, ,  \label{eq:dgammaSol}
\eeq
where the first term is the homogeneous solution with constant superhorizon initial condition~$d_{\gamma,\in} \equiv d_\gamma(y_\in \ll 1)$.  The second term is the inhomogeneous correction induced by the evolution of metric fluctuations. Since $\sin(y- y') = \sin y \cos y' - \cos y \sin y'$, we can write (\ref{eq:dgammaSol})~as
\beq
d_\gamma(y) = \Big[d_{\gamma,\in} + c_\gamma^{-2} A(y) \Big] \cos y - c_\gamma^{-2} B(y) \sin y \, , \label{eq:dg2}
\eeq
where 
\begin{align}
A(y) &\equiv  \int^y_0 \d y'\, \Phi_+(y')\hskip 1pt\sin y'\, ,	\label{eq:A}	\\[4pt]
B(y) &\equiv  \int^y_0  \d y'\, \Phi_+(y')\hskip 1pt\cos y'	\, .  \label{eq:B}
\end{align} 
We wish to evaluate (\ref{eq:dg2}) at recombination, $y \to y_\rec$.  For the high-$\ell$ modes of the CMB, it is a good approximation to take the limit $y\to \infty$ and assume that the background is radiation dominated (see Appendix~\ref{app:ext} for further discussion). If the integral in (\ref{eq:B}) converges in this limit, then a non-zero value of $B \equiv \lim_{y \to \infty} B(y)$ will produce a constant phase shift $\theta$ relative to the homogeneous solution,
\beq
\sin\theta = \frac{B}{\sqrt{\left(A + c_\gamma^2\hskip 1pt d_{\gamma,\in}\right)^2+B^2}}\, . \label{eq:theta}
\eeq
This phase shift will be reflected in a shift of the acoustic peaks of the CMB anisotropy spectrum.  In the following, we will identify the precise physical conditions for which such a phase shift is generated.

\vskip 4pt			
It will be convenient to combine $B$ and $A$ into a complex field 
\beq
B+iA \,=\, \int_0^\infty \d y\, e^{i y}\, \Phi_+(y)  \,=\, \frac{1}{2} \int_{-\infty}^\infty \d y\, e^{i y} \left[  \Phi^{(S)}_+(y) + \Phi^{(A)}_+(y)\right]\, ,  \label{eq:AB}
\eeq
where $\Phi^{(S)}_+(y)$ is an even function of $y$, while $\Phi^{(A)}_+(y)$ is an odd function.  It is easy to see that the even part of $\Phi_+$ determines $B$ and the odd part determines $A$:
\beq
B =  \frac{1}{2} \int_{-\infty}^\infty \d y\, e^{i y} \, \Phi^{(S)}_+(y)\, , \qquad i A = \frac{1}{2}   \int_{-\infty}^\infty \d y\, e^{i y} \, \Phi^{(A)}_+(y) \, . \label{equ:Bdef}
\eeq
We will get $B=0$ as long as $\Phi^{(S)}_+(y)$ is an analytic function and $e^{i y} \Phi^{(S)}_+(y)$ vanishes faster than $y^{-1}$ for $|y| \to \infty$.\footnote{Since the equations are symmetric in $y \to -y$, the odd part $ \Phi^{(A)}_+(y)$ is not analytic around $y= 0$. This is why we always find contributions to $A$.}  This suggests two ways of generating a non-zero $B$ and, hence, a phase shift in the solution for the photon density:
\vskip 8pt
\begin{center}
\begin{tabular}{r l c l}
{\it i.} & rapid growth of $\Phi_+^{(S)}(\pm iy)$ & $\longrightarrow$ & mode traveling faster than $c_\gamma$, \\[4pt]
{\it ii.} & non-analytic behavior of $\Phi^{(S)}_+(y)$ &$\longrightarrow$ & non-adiabatic fluctuations.
\end{tabular}
\end{center} 
\vskip 8pt
The mathematical requirements listed on the left are mapped directly into physical conditions, shown on the right.  

\begin{itemize}
\item The first condition is easy to understand physically: in~(\ref{eq:AB}), the Green's function of $d_\gamma$, i.e.~$\sin(y-y')$, leads to exponential suppression for $y \to i \infty$.  To have a growing solution at $y = i \infty$, we therefore need a term in $\Phi_+$ of the form $e^{-i c_s k \tau} = e^{-i(c_s/c_\gamma)y}$ with $c_s > c_\gamma$.\footnote{Note that $c_s$ is just a parameter of the wave-like solution and is not necessarily the sound speed of a fluid.  Indeed, in the case of free-streaming radiation, it corresponds to the propagation speed of the individual particles.}
\item The second possibility, non-analyticity, is easy to understand mathematically, but the physical requirements are less transparent.  First of all, the equations of motion for any mode should be analytic around any finite value of $k \tau$ in the radiation-dominated era since there is no preferred time.  Hence, the only moment at which non-analytic behavior is possible is around $k\tau = 0$, i.e.~where the initial conditions are defined.  Let us first show that adiabatic initial conditions are analytic at $k\tau = 0$.  By definition,  for adiabatic initial conditions, any long-wavelength mode is locally generated by a diffeomorphism~\cite{Weinberg:2003sw}.  In the limit $k \tau \to 0$, we then have $\Phi_+ = \Phi_{+,\in} + {\cal O}(k^2 \tau^2)$.  This expansion is necessarily analytic in $k^2$ (by locality and rotational invariance), but also in $k^2\tau^2$, because the scaling $k \to \lambda k$ and $ \tau  \to \lambda^{-1} \tau$ can be absorbed into the overall normalization of the scale factor $a$ which has no physical effect.\footnote{In a universe with a preferred time, this rescaling would also require a shift in this preferred time to keep the density fluctuations fixed.  For adiabatic modes, the curvature perturbation $\zeta$ is conserved outside the horizon even in the presence of such a preferred time and so this is unlikely to have an impact on gauge-invariant observables.} Hence, $\Phi^{(S)}_+(y)$ must be analytic around $y = c_\gamma k \tau = 0$, as long as the modes are adiabatic.  Conversely, any violation of analyticity requires a source of non-adiabaticity.
\end{itemize}
In the following sections, we will illustrate the different physical origins of the CMB phase shift through a number of simple examples.

\subsection{Intuition from a Toy Model}
\label{sec:toyModel}

To gain more intuition for the system of equations discussed in the previous section, let us solve them exactly in a simple toy model. In particular, we will study an example in which the metric fluctuations $\Phi_+$ propagate with a different speed than the photons, $c_s \ne c_\gamma$. We wish to understand under which conditions this mismatch leads to a phase shift in the photon oscillations.

\vskip 4pt
In the absence of anisotropic stress, the Einstein equation for $\Phi_+$, eq.~(\ref{eq:Einstein0}), is
\beq
\ddot \Phi_+ + 3 \H \dot \Phi_+ + (2\dot \H + \H^2) \Phi_+ \,=\, 8\pi G a^2\, \delta P\, . \label{eq:Einstein}
\eeq
The Friedmann equations describing the evolution of the homogeneous background imply
\begin{align}
2\dot \H + \H^2 &= - 8\pi G a^2\, \bar P= - 3\H^2 w\, , \label{eq:H}
\end{align}
where we have defined the equation of state $w \equiv \bar P/\bar \rho$. (Note that we are {\it not} assuming that $P=w\rho$.) We write the pressure perturbation as
\beq
\delta P = c_s^2\hskip 1pt \delta \rho + \delta P_\mathrm{en}\, , \label{eq:deltaP}
\eeq
where $c_s$ is the speed controlling the propagation of the total density perturbation $\delta \rho$ and $\delta P_\mathrm{en}$ denotes the non-adiabatic entropy perturbation. For adiabatic fluctuations, one has $\delta P_\mathrm{en} = 0$ and $c_s^2 = w- [3\mathcal{H}(1+w)]^{-1}\hskip 2pt \dot w$.  We  eliminate the density perturbation $\delta \rho$ using the relativistic generalization of the Poisson equation (\ref{eq:Poisson}).  Equation~(\ref{eq:Einstein}) can then be written as
\beq
\ddot \Phi_+ + 3 \H (1+c_s^2) \dot \Phi_+ - 3 \H^2 (w- c_s^2) \Phi_+ + c_s^2 k^2 \Phi_+ \,=\, 8\pi G a^2\, \delta P_\mathrm{en}\, . \label{eq:Phi++0}
\eeq

So far, this is fairly general and has only assumed vanishing anisotropic stress.  In particular, at this point $w$ and $c_s$ are still general, possibly time-dependent parameters.  To be able to derive an analytic solution for the evolution of $\Phi_+(\tau)$, we will now make a few simplifying assumptions. First, we assume that the equation of state $w$ is nearly constant,  so that we can integrate~(\ref{eq:H}) to get
\beq
\H =  \frac{2}{1+3w} \frac{1}{\tau}\, .
\eeq
Second, we take $c_s^2 \approx \mathrm{const.}$ and $\delta P_\mathrm{en} \approx 0$.  This allows us to solve (\ref{eq:Phi++0}) analytically.  For arbitrary $w$ and $c_s$, these assumptions are not guaranteed to be easily realizable in a physical model.  Our analysis only serves as a simple illustration of some of the effects that give rise to phase shifts in the CMB.  More concrete examples of these effects will be discussed in \textsection\ref{sec:iso} and~\textsection\ref{sec:fs}.

\vskip 4pt
It is convenient to define $z \equiv \c k \tau$ and write~(\ref{eq:Phi++0}) as
\beq
\frac{\d^2}{\d z^2}\Phi_+ +  \frac{1-2\alpha}{z} \, \frac{\d}{\d z}\Phi_+  +\left[1-  \frac{\beta}{z^2} \right]\Phi_+  =  0\, ,	\label{eq:Phi++}
\eeq
with
\beq
\alpha \equiv \frac{1}{2}-\frac{3(1+\c^2)}{1+3w}\ , \qquad\ \beta \equiv \frac{12 (w-\c^2)}{(1+3w)^2}\, . \label{eq:abG}
\eeq	
In the physically interesting parameter regime, $0 \leq (c_s^2, w) \leq 1$,  we have $-\frac{11}{2} \leq \alpha \leq -\frac{1}{4}$, with equality for $(c_s^2,w) = (1,0)$ and $(0,1)$, respectively.  The general solution of~\eqref{eq:Phi++} is 
\beq
\Phi_+(z) = z^{\alpha}\left(c_1J _{\kappa}(z)+ c_2Y _{\kappa}(z)\right)\, , \qquad \kappa \equiv \sqrt{\alpha^2+\beta}\, ,	\label{eq:Phi+nfs}
\eeq
where $J _{\kappa}(z)$ and $Y _{\kappa}(z)$ are Bessel functions of the first and second kind, respectively.  Note that $\kappa$ is strictly positive and real-valued for the physically relevant parameter regime: $ \frac{1}{2}\sqrt{3} \leq \kappa \leq \frac{1}{2} \sqrt{73} $, with the minimum at $(c_s^2,w) = (\frac{1}{3},1)$ and the maximum at $(c_s^2,w) = (1,0)$. 

\begin{figure}[h!t]
\begin{center}
\includegraphics[scale=1.0]{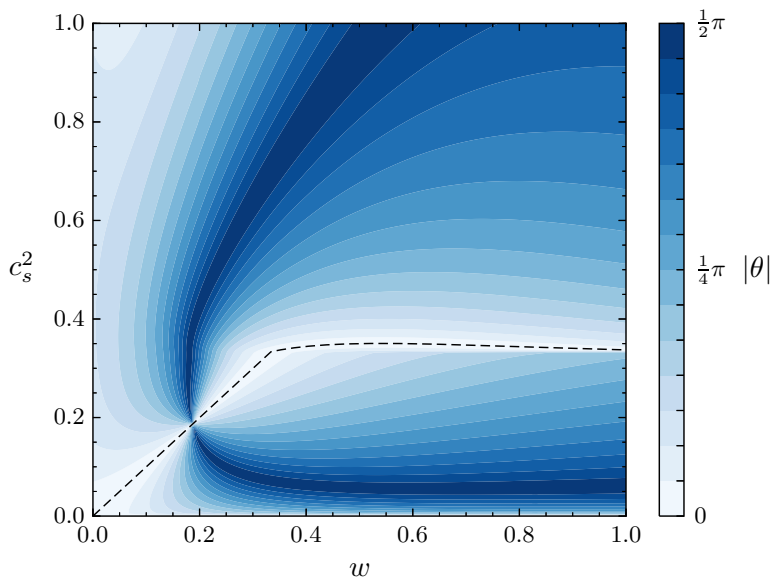}
\caption{Phase shift $\theta$ for varying speed of sound ($\c$) and equation of state ($w$). The dashed line denotes~$\theta=0$. Below this line, the phase shift is negative, while above it is positive.}
\label{fig:toyModel_theta_contour}
\end{center}
\end{figure}

\vskip 4pt
To impose initial conditions, we consider the superhorizon limit, $z \ll 1$, 
\beq
\Phi_+(z) \simeq \frac{2^{-\kappa} \left(c_1+c_2 \cot{(\pi\kappa)}\right)}{\Gamma(1+\kappa)} z^{\alpha +\kappa} - \frac{2^{\kappa} c_2 \Gamma(\kappa)}{\pi} z^{\alpha-\kappa} + \cdots\, .	\label{eq:Phi+Superhorizon}
\eeq
Since $\alpha + \kappa > \alpha-\kappa$, the ``growing mode'' solution corresponds to $c_2 \equiv 0$.  Hence, we have
\beq
\Phi_+(z) = c_1 z^{\alpha} J _{\kappa}(z)\, .   \label{eq:Phi+nfs2}
\eeq
The overall normalization in \eqref{eq:Phi+nfs2} will depend on the nature of the initial conditions (adiabatic or entropic). For $c_s^2 = w$, the superhorizon limit of $\Phi_+$ is a constant, which we match to the superhorizon value of the primordial curvature perturbation $\zeta$. This leads to the normalization $c_1 =2 \sqrt{2 \pi}\,\zeta$. We will maintain this normalization even for $c_s^2 \ne w$, although, in principle, the normalization of non-adiabatic modes is model-dependent.  The $y \to \infty$ limit of~\eqref{eq:B} then becomes
\beq
B = 2 \sqrt{2 \pi}\,\zeta \int^\infty_0 \d y\, \left(\frac{\c}{c_\gamma} y\right)^{\!\alpha}  J_\kappa\!\left(\frac{\c}{c_\gamma} y\right)\cos y\, , \label{eq:BB}
\eeq
and similarly for $A$. Together with $d_{\gamma,\in}=-3\zeta$, this allows us to compute the phase shift via~(\ref{eq:theta}).  A graphical illustration of the dependence of the phase shift $\theta$ on the parameters $\c^2$ and $w$ is given in Fig.~\ref{fig:toyModel_theta_contour}.  Let us emphasize again that we do not imagine that all of the combinations of $\c^2$ and $w$ that we show in the figure can be realized in a physically realistic model. In the following, we take slices through the parameter space to show that the most important features of the figure can be understood analytically.

\begin{figure}[h!t]
\begin{center}
\includegraphics[scale=1.0]{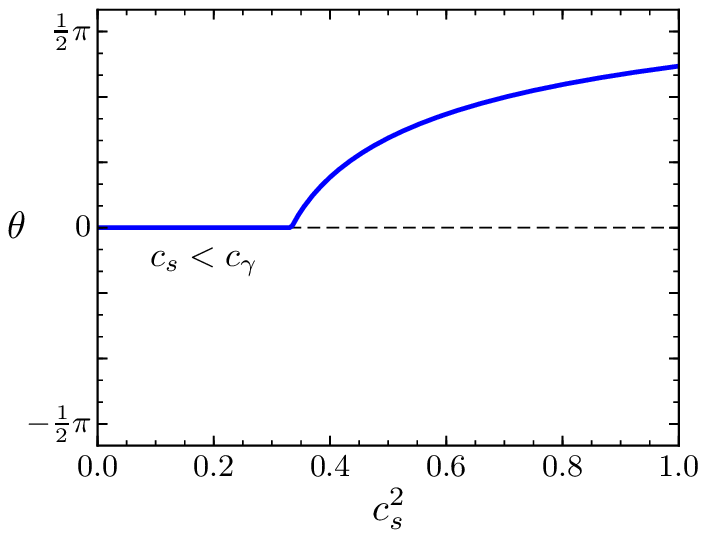} \hspace{0.5cm}
\includegraphics[scale=1.0]{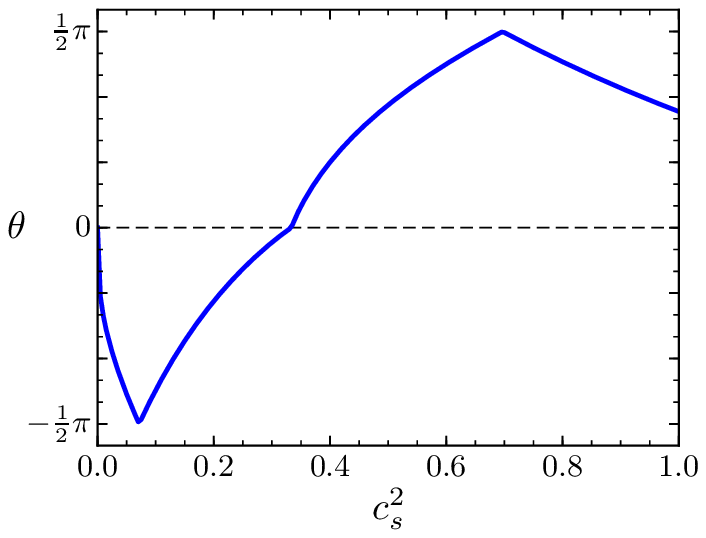}
\vskip -8pt
\caption{Phase shift $\theta$ for varying $c_s^2=w$ (\textit{left}) and for varying $c_s^2$ and fixed $w=\frac{1}{3}$ (\textit{right}).}
\label{fig:toyModel_theta_cases}
\end{center}
\end{figure}

\vskip 4pt
Consider first the special case $c_s^2 = w$, which corresponds to adiabatic fluctuations.  The parameters in~(\ref{eq:abG}) and~(\ref{eq:Phi+nfs}) then reduce to
\beq
\alpha = - \frac{5+3\c^2}{2(1+3\c^2)}\, , \qquad \beta =0\, , \qquad \kappa = |\alpha| \, .
\eeq		
As shown in the left panel of Fig.~\ref{fig:toyModel_theta_cases}, the phase shift then vanishes for $\c \leq c_\gamma = \frac{1}{\sqrt{3}}$ and is positive for $\c > c_\gamma$. This is consistent with our abstract reasoning of the previous section. At large $z= c_s k\tau$, the solution (\ref{eq:Phi+nfs2}) behaves as $z^{\alpha - \frac{1}{2}} \cos(z) \propto \cos(\c/c_\gamma\, y)$, with $y=c_\gamma k \tau$.  Since the contour at infinity in~(\ref{equ:Bdef}) will not vanish when $\c > c_\gamma$, we cannot conclude that $\theta =0$ (cf.~condition \textit{i.}\ in \textsection\ref{sec:originPhaseShift}).  To find $\theta \neq 0$ was still not guaranteed, but there was no reason to expect otherwise.  In contrast, $\theta$ vanishes for $\c \leq c_\gamma $ for exactly the reasons discussed before. In particular, the solution~\eqref{eq:Phi+nfs2} now takes the form $z^\alpha  J_{|\alpha|} (z)$ with $\alpha < 0$.  Near $z=0$, the solution is analytic (cf.~condition \textit{ii.}\ in \textsection\ref{sec:originPhaseShift}) since the non-analytic behavior of $z^\alpha$ cancels that of the Bessel function.  Of course, this is precisely what we expected for adiabatic modes.

\vskip 4pt
Taking $\c^2 \neq w$ corresponds to non-adiabatic fluctuations, i.e.~fluctuations which contain an isocurvature component.  In this case, we expect a phase shift to arise for any values of $\c^2$ and $w$. To illustrate this, let us take $w = \frac{1}{3}$ and only allow $\c^2$ to vary. We then have
\beq
\alpha = -1 -\frac{3\c^2}{2}\, , \qquad  \beta = 1-3\c^2\, , \qquad \kappa = \frac{1}{2}\sqrt{8+9\c^4}\, .
\eeq
The corresponding phase shift is shown in the right panel of~Fig.~\ref{fig:toyModel_theta_cases}.  We see that the phase shift now only vanishes at the special point $\c^2 = w = \frac{1}{3}$, where the fluctuations are adiabatic. This is also where the parameter $\beta$ changes sign, which is the origin of the change in the sign of the phase shift, $\theta \lessgtr 0$ for $\c^2 \lessgtr w$.
This time the phase shift is associated with the non-analytic behavior of $\Phi_+(z)$ near the origin.  To see this explicitly, consider the $z \to 0$ limit of (\ref{eq:Phi+nfs2}):
\beq
\Phi_+(z) = \frac{c_1}{2^\kappa \Gamma(1+\kappa)}  \, z^{\alpha + \kappa} \left[1+{\cal O}(z^2)\right] \, . \label{eq:Phi+z}
\eeq
For $0\leq c_s^2 \leq 1$, we have $\alpha+\kappa < 2$ and, hence, $\Phi_+(z)$ is non-analytic at $z=0$ (cf.~condition \textit{ii.}\ in \textsection\ref{sec:originPhaseShift}), except for the special case of the adiabatic limit where $\alpha+\kappa = 0$. This corresponds to the non-trivial superhorizon evolution of $\Phi_+$ in the presence of isocurvature modes.

\subsection{Simple Examples}

\subsubsection{Isocurvature Perturbations}
\label{sec:iso}

The toy model of the previous section suggests that isocurvature perturbations produce a phase shift. To study this further, it is useful to consider a slightly more realistic model.  To simplify the calculations as much as possible, our curvaton-like model will include three species:  photons~($\gamma$), a dark radiation fluid ($Y$) and a matter component ($m$) that decays into the dark radiation.  The matter will carry the isocurvature fluctuations. We include the dark radiation because we are only interested in the gravitational effects on the photons, i.e.~we want to avoid the direct heating of the photons due to the decaying matter.  The model will allow us to explore small deviations from the adiabatic limit $c_s^2=w$.

\vskip 4pt
The coupled equations for the background densities of $m$ and $Y$ are 
\begin{align}
\frac{1}{a^3} \frac{\d}{\d\tau}(a^3 \bar \rho_m) &= - \Gamma \hskip 1pt a  \hskip 2pt \rho_m \, , \label{eq:rhom}\\[4pt]
\frac{1}{a^4} \frac{\d}{\d\tau}(a^4 \bar \rho_Y)  &= + \Gamma \hskip 1pt a  \hskip 2pt \rho_m \, , \label{eq:rhoY}
\end{align}
where $\Gamma$ is a constant decay rate. To simplify calculations, we will work perturbatively in $\epsilon_m \equiv \bar \rho_m / \bar \rho$.  At zeroth order in $\epsilon_m$, the universe is radiation dominated, and hence $a = \tau / \tau_\in$. Integrating (\ref{eq:rhom}), we get
\beq
\bar\rho_m(a) = \frac{\bar\rho_{m,\in}}{a^3} \,e^{-\frac{1}{2}\Gamma \tau_\in(a^2-1)} \, , \label{eq:rhom2}
\eeq
where we set the initial value $\bar\rho_{m,\in} \equiv \bar\rho_m(\tau_\in)$. Substituting (\ref{eq:rhom2}) into (\ref{eq:rhoY}), we would get the solution for $\bar\rho_Y(a)$, however, this will not be needed for our purposes. 

\vskip 4pt
We now wish to determine how the decaying matter affects the evolution of the metric perturbation $\Phi_+$. In the absence of anisotropic stress, this is given by~(\ref{eq:Einstein}).  The pressure perturbations only receive contributions from $\gamma$ and $Y$, so we have 
\beq
\delta P = c_\gamma^2(\delta \rho_\gamma + \delta \rho_Y) \,=\, \frac{1}{3}\delta \rho - \frac{1}{3}\delta \rho_m\, ,
\eeq
where we have used $c_Y^2= c_\gamma^2 = \frac{1}{3}$.  Using the Poisson equation (\ref{eq:Poisson}), we can write (\ref{eq:Einstein}) as
\beq
\ddot \Phi_+ + 4 \H \dot \Phi_+ - \frac{1}{3} \nabla^2 \Phi_+ \,=\,  (3w - 1) \H^2 \Phi_+ - \H^2  \epsilon_m \delta_m \, . \label{PhiPlus}
\eeq
We wish to solve this at linear order in $\epsilon_m$.

\vskip 4pt
We will shortcut the computation by isolating the isocurvature contribution.  Suppose we write $\Phi_+ = \Phi_+^{\rm ad} + \Phi_+^{\rm iso}$, and similarly for $\delta_m$.  Equation~(\ref{PhiPlus}) then implies
\beq
\Phi_+^{\rm iso} \hskip 1pt {}' {}'  + \frac{4}{y}  \Phi_+^{\rm iso} \hskip 1pt  {}' +  \Phi_+^{\rm iso} \,=\,-\frac{\epsilon_m}{y^2}\hskip 1pt  \delta_m^{\rm iso} + {\cal O}(\epsilon_m^2) \, , \label{eq:Phi+2}
\eeq
where we have used that $\Phi_+^{\rm iso} \sim {\cal O}(\epsilon_m)$, so that all terms multiplying $\Phi_+^{\rm iso}$ can be evaluated at zeroth order in $\epsilon_m$. Since the right-hand side of (\ref{eq:Phi+2}) is proportional to $\epsilon_m$, we only need the homogeneous solution for $\delta_m^{\rm iso}$, which is
\beq
\delta_m^{\rm iso}(y) = c_1 + c_2 \ln y\, , \label{equ:deltaIso}
\eeq
where $c_{1,2}$ are constants that may depend on $k$. We solve~(\ref{eq:Phi+2}) using the Green's function~(\ref{eq:Gplus}). Substituting~(\ref{eq:rhom2}) and~(\ref{equ:deltaIso}), we get		
\beq
\Phi_+^{\rm iso}(y) \,=\, \frac{1}{c_\gamma k \tau_\in}\, \tilde \epsilon_{m,\in}\, \underbrace{\int_{y_\in}^y \d y'\, G_{\Phi_+}(y,y') \, \left(-\exp\left[-\frac{1}{2}\frac{(y')^2}{(c_\gamma k \tau_{\rm dec})^2}\right] \, \frac{c_1 + c_2 \ln(y')}{y'} \right)}_{\displaystyle \equiv  {\cal I}(y)}\, , \label{eq:PhiPlusF}
\eeq
where $\tilde \epsilon_{m,\in} \equiv \epsilon_{m,\in} \hskip 1pt e^{\frac{1}{2}(\tau_\in/\tau_{\rm dec})^2}$ and we have introduced the ``decay time scale'' $\tau_{\rm dec}^2 \equiv \tau_\in/\Gamma$.

Let us comment on a few features of this solution.  First of all, we notice that the integral is highly suppressed when $k \tau_{\rm dec} \ll 1$.  The reason is easy to understand: the integral would have been dominated by contributions around the time of horizon crossing, $y \sim {\cal O}(1)$, but, for $k \tau_{\rm dec} \ll 1$, this is long after $\rho_m$ has decayed.  Second, we see that the solution has an overall factor of $(c_\gamma k \tau_\in)^{-1}$.  This reflects the growth of $\epsilon_m$ from the initial time, $\tau_\in$, to the time of horizon crossing, $(c_\gamma k)^{-1}$.

\begin{figure}[ht!]
\begin{center}
\includegraphics[scale=1.0]{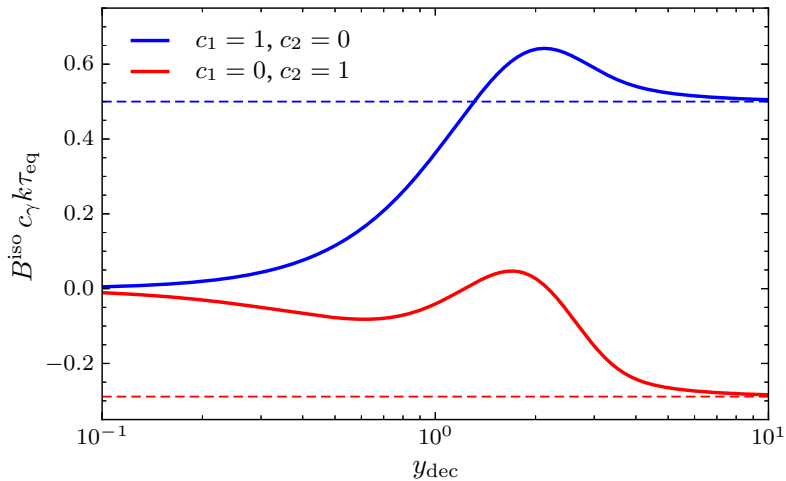} 
\vskip -8pt
\caption{Numerical value of $B^{\rm iso} \, c_\gamma k \tau_{\rm eq}$ as a function of $y_{\rm dec}$.  The blue and red solid lines show the effect from $c_1$ and $c_2$, respectively.  The dashed lines are the asymptotic values  calculated in~(\ref{eq:Bisofinal}).}
\label{fig:Isocurvature}
\end{center}
\end{figure}

\vskip 4pt
It is convenient to define $\tau_{\rm eq}$ as the time at which $\bar \rho_m$ and $\bar \rho_\gamma + \bar \rho_Y$ would be equal \textit{if} there was no decay. This is given by $\tau_{\rm eq} \simeq \tau_\in/\tilde\epsilon_{m,\in}$. Equation~(\ref{eq:PhiPlusF}) then becomes 
\beq
\Phi_+^{\rm iso}(y) =  \frac{1}{c_\gamma k \tau_{\rm eq}}\, {\cal I}(y) \, .	\label{eq:PhiPlusF2}
\eeq
We compute the phase shift by substituting~(\ref{eq:PhiPlusF2}) into~(\ref{eq:A}) and~(\ref{eq:B}), and taking the limits $y_\in \to 0$ and $y \to \infty$,
\begin{align}
A^{\rm iso} &\equiv   \frac{1}{c_\gamma k \tau_{\rm eq}} \int^\infty_0 \d y'\, {\cal I}(y')\hskip 1pt\sin y'\, ,	\label{eq:A2}	\\[4pt]
B^{\rm iso} &\equiv  \frac{1}{c_\gamma k \tau_{\rm eq}} \int^\infty_0  \d y'\, {\cal I}(y')\hskip 1pt\cos y'		\, . \label{eq:B2}
\end{align}
In Fig.~\ref{fig:Isocurvature} we display the numerical result for $B^{\rm iso}$ as a function of $y_{\rm dec} \equiv c_\gamma k \tau_{\rm dec}$.  In the limit $y_{\rm dec} \gg 1$, we can simplify the calculation by dropping the exponential in (\ref{eq:PhiPlusF}).  We then get
\begin{align}
A^{\rm iso} &\,\xrightarrow{\ y_{\rm dec} \gg 1 \ }\, \frac{\pi}{4} c_2\,  \frac{1}{c_\gamma k \tau_{\rm eq}}\, , \\
B^{\rm iso} &\,\xrightarrow{\ y_{\rm dec} \gg 1 \ }\, \frac{1}{2}( c_1 - c_2 \gamma_{\rm E} )\,  \frac{1}{c_\gamma k \tau_{\rm eq}} \, ,  \label{eq:Bisofinal}
\end{align}
where $\gamma_{\rm E} \approx 0.5772$ is the Euler-Mascheroni constant.  We see from Fig.~\ref{fig:Isocurvature} that the analytic result~(\ref{eq:Bisofinal}) becomes a good approximation for $y_{\rm dec} \gtrsim 5$. 

To summarize, we have demonstrated in a simple model that isocurvature perturbations give rise to a phase shift, as we expected from condition \textit{ii.}~of \textsection\ref{sec:originPhaseShift}. As suggested by Fig.~\ref{fig:Isocurvature} this phase shift has a nontrivial scale dependence which probably allows it to be distinguished from other sources for a phase shift. This scale dependence is likely to be a general feature of isocurvature models.

\subsubsection{Free-Streaming Particles}
\label{sec:fs}
 
Above we have seen that a phase shift is also generated if fluctuations in the gravitational potential travel faster than the speed of sound in the photon-baryon fluid. A simple way to realized this is through free-streaming relativistic particles,\hskip1pt\footnote{While it should be physically clear that free-streaming radiation travels at the speed of light, this property is sometimes not very transparent in the equations for the density perturbations of this radiation.  Instead, what is usually more apparent is that free-streaming particles can induce a significant anisotropic stress, which then provides a source for $\Phi_+$ and, hence, the evolution of $d_\gamma$ through (\ref{eq:dgamma}).  The origin of the phase shift is therefore often identified with the presence of anisotropic stress.  However, in principle, one could imagine situations with significant anisotropic stress, but no supersonic propagation modes (e.g.~non-relativistic, free-streaming particles). In those cases, we would not expect a phase shift. Hence, it is the propagation speed, not the anisotropic stress itself, that makes the phase shift possible.}
such as neutrinos~\cite{Bashinsky:2003tk}.  In this section, we give a new derivation of this result.  In Section~\ref{sec:analysis}, we will show that the CMB data is now accurate enough to detect this effect (see also~\cite{Follin:2015hya}).
 
\vskip 4pt
Since most of the modes relevant to current and future CMB observations entered the horizon during the era of radiation domination, our discussion in this section will ignore both the matter and baryon content of the universe.  In Appendix~\ref{app:ext}, we show that this is a good approximation for high-$\ell$ modes and also discuss some of the implications of a finite matter density. 

\vskip 4pt
We start with the Einstein equation during radiation domination, eq.~(\ref{eq:Phi+RD}), which in the absence of additional pressure perturbations $\delta P_a = c_a^2 \delta\rho$, with $c_a \neq c_\gamma$,  takes the form
\begin{align}
\Phi_+'' + \frac{4}{y} \Phi_+' + \Phi_+ &\,=\, \tilde {\cal S}[\Phi_-]  \nonumber \\
&\,\equiv\, \Phi_-'' + \frac{2}{y} \Phi_-'  + 3\Phi_-  \, . \label{eq:EE}
\end{align}
Using the Green's function (\ref{eq:Gplus}), the solution for $\Phi_+$ can be written as
\beq
\Phi_+(y) = 3\Phi_{+,\in}\, \frac{\sin y - y \cos y}{y^3}  + \int_{y_\in}^y  \d  y' \, \tilde {\cal S}[\Phi_-(y')]\,G_{\Phi_+}(y, y')\, . \label{eq:Phi+}
\eeq	
Following~\cite{Bashinsky:2003tk}, we will write this as an expansion in powers of the fractional energy density contributed by the species of free-streaming particles, $X$, as measured by the dimensionless ratio
\beq
\epsilon_X \equiv \frac{\rho_X}{\rho_{\gamma} + \rho_{X}}  \, = \, \frac{\Nf}{a_\nu + \Nf} \, .
\eeq
For the Standard Model neutrinos, we have $\epsilon_\nu \approx 0.41$. We determine the superhorizon initial condition of the homogeneous solution, $\Phi_{+,\in}$, by matching to the constant superhorizon solution for adiabatic initial conditions~\cite{Bashinsky:2003tk},
\beq
\Phi_{+,\in} = \frac{20 + 4 \epsilon_X}{15 + 4 \epsilon_X}\,\zeta = \frac{4}{3}  \zeta \left( 1 - \frac{1}{15}\epsilon_X + {\cal O}(\epsilon_X^2) \right) \, ,
\eeq
where $\zeta$ is the conserved curvature perturbation.

To find the inhomogeneous part of the solution (\ref{eq:Phi+}), we require $\Phi_-(y)$. This is determined by the anisotropic stress $\sigma_X$ induced by the free-streaming particles,
\beq
\Phi_-(y) = -\frac{2 k^2 \epsilon_X}{y^2} \sigma_X(y) =  - \frac{4}{3}\frac{\epsilon_X}{y^2} D_{X,2}(y)\, , \label{eq:Phi-X}
\eeq
where $D_{X,2}$ is the quadrupole moment of the momentum-integrated distribution function defined in~(\ref{eq:distribution}). The first equality follows from the Einstein constraint equation~\eqref{eq:EinsteinC} with $\sigma = \epsilon_X \sigma_X$, whereas the second equality employs~\eqref{eq:DmultipoleLinks}. The evolution of $D_{X,2}$ is determined from the linearized, collisionless Boltzmann equation~(\ref{eq:Boltzmann}),
\beq
\dot D_X + i k \mu D_X = - 3 i k \mu\, \Phi_+ \, . \label{eq:BoltzmannDX}
\eeq
Defining $D_{X,\in} \equiv D_X(\tau_\in)$ at some time $\tau_\in$, the solution to~\eqref{eq:BoltzmannDX} is
\beq
D_X(\tau) = e^{- i k \mu (\tau-\tau_\in)} D_{X,\in}- 3 i k \mu  \int_{\tau_\in}^\tau \d \tau'\, e^{- i k\mu (\tau - \tau')} \Phi_+(\tau') \, . \label{eq:DX}
\eeq
We wish to extract the quadruple moment $D_{X,2}$ of the solution.  Since $D_{X,\ell}(\tau_\in) \propto \tau_\in^\ell$, we will only keep the monopole term $D_{X,0}(\tau_\in)$ in the homogeneous part of the solution. This is possible because we can set the initial conditions at a sufficiently early time $\tau_\in \ll k^{-1}$, so that the modes with $\ell>0$ will be subdominant. In fact, we will take $k\tau_\in \to 0$ from now on.  Assuming adiabatic initial conditions, i.e.~$D_{X,0}(\tau_\in) =d_{X,\in}=-3\zeta$, we get
\beq
D_{X,2}(y) = -3\zeta\, j_{2}\!\left[c_\gamma^{-1}y\right] + \frac{3}{c_\gamma} \int_0^y \d y'\, \Phi_+(y') \left\{ \frac{2}{5} j_{1}\!\left[c_\gamma^{-1}(y-y')\right] - \frac{3}{5}j_{3}\!\left[c_\gamma^{-1}(y-y')\right]  \right\} \, ,	\label{eq:DX2}
\eeq
where the Bessel functions $j_\ell$ arise from the Rayleigh expansion of the exponentials.  Substituting this into~(\ref{eq:Phi-X}) directly links the two gravitational potentials $\Phi_+$ and $\Phi_-$.  The most important feature of the solution~(\ref{eq:DX2}) is that it contains modes that travel at the speed of light.  Specifically, recall that $c_\gamma^{-1} y = k \tau$ and, therefore, the Bessel functions describe oscillatory solutions with a speed of propagation of $c_s = 1$.  As we have emphasized before, this is the property of the free-streaming radiation that makes a phase shift possible.

\vskip 6pt
The above is a closed set of equations which we can solve perturbatively in $\epsilon_X$:
\beq
\Phi_\pm \equiv \sum_{n} \Phi_\pm^{(n)}\, , \quad \ d_\gamma \equiv \sum_n d_\gamma^{(n)}\, ,
\eeq
where the superscripts on $ \Phi_\pm^{(n)}$ and $d_\gamma^{(n)}$ count the order in $\epsilon_X$.  Here, we present the solution up to first order:

\begin{itemize}
\item At zeroth order in $\epsilon_X$, we have $\Phi_-^{(0)}(y) = 0$ and, hence, $\Phi_+^{(0)}$ is given by the homogeneous solution,
\beq
\Phi_+^{(0)}(y) = 4\zeta\, \frac{\sin y - y \cos y}{y^3}\, .	\label{Phi+h}
\eeq
Inserting this into~(\ref{eq:A}) and~(\ref{eq:B}), we find
\begin{align}
A^{(0)}(y) &\,=\, 2\zeta - 2\zeta\, \frac{\sin^2(y)}{y^2} \hskip 12pt \xrightarrow{\, y\to\infty\, }\ 2\zeta \, , \label{eq:A0} \\
B^{(0)}(y) &\,=\, 2\zeta\, \frac{y - \cos y \sin y}{y^2}	\ \xrightarrow{\, y\to\infty\, }\ 0 \, . \label{eq:B0}
\end{align}
The result for the photon density perturbations then is
\beq
d_\gamma^{(0)}(y) \approx  3\zeta \cos y	\, . \label{eq:d0}
\eeq
We conclude that in the absence of anisotropic stress, the correction due to $\Phi_+$ is in phase with the homogeneous solution and $B$ vanishes as expected. 

\item At first order in $\epsilon_X$, we only need the zeroth-order solution of the anisotropic stress, $\sigma_X^{(0)}$, since the source in~\eqref{eq:Phi-X} already comes with an overall factor of $\epsilon_X$. Hence, eqs.~\eqref{eq:Phi-X} and~\eqref{eq:DX2} can be written as
\begin{align}
\Phi_-^{(1)}(y)	&\,=\, 4 \zeta\,\frac{\epsilon_X}{y^2}\, j_{2}\!\left[c_\gamma^{-1}y\right] \label{eq:Phi-X2}	\\[2pt]
		& \ \quad - \frac{4 }{c_\gamma}  \frac{\epsilon_X}{y^2} \int_0^y \d y'\, \Phi_+^{(0)}(y') \left\{ \frac{2}{5} j_{1}\!\left[c_\gamma^{-1}(y-y')\right] - \frac{3}{5}j_{3}\!\left[c_\gamma^{-1}(y-y')\right]  \right\} \, , \nonumber
\end{align}
where $\Phi_+^{(0)}$ is given by (\ref{Phi+h}).  Substituting (\ref{eq:Phi-X2}) into~(\ref{eq:Phi+}), we obtain
\beq
\hskip -10pt \Phi_+^{(1)}(y) = - \frac{4}{15} \zeta\, \epsilon_X\, \frac{\sin y - y \cos y}{y^3}  + \int_0^y  \d  y' \, \tilde {\cal S}[\Phi_-^{(1)}(y')]\,G_{\Phi_+}(y, y') \, .
\label{eq:Phi+R}
\eeq
Inserting this into~(\ref{eq:A}) and~(\ref{eq:B}), we finally get expressions for $A^{(1)}$ and $B^{(1)}$.  These have to be evaluated numerically, and we find
\beq
A^{(1)} \approx - 0.268\, \zeta\, \epsilon_X\, ,	\qquad	B^{(1)} \approx 0.600\, \zeta\, \epsilon_X\, .\label{eq:A+B_numericsCcodeX}
\eeq
The non-zero value of $B^{(1)}$ corresponds to the expected phase shift. 
\end{itemize}
Using~(\ref{eq:theta}) with  $d_{\gamma,\in}=-3\zeta$, we get 
\beq
\theta \approx 0.191\pi\, \epsilon_X + \mathcal{O}(\epsilon_X^2)\, , \label{eq:thetaX}
\eeq
which is consistent with the result of Bashinsky\,\&\,Seljak~\cite{Bashinsky:2003tk}.

\vskip 4pt
The phase shift is a clean signature of free-streaming particles and will naturally play an important role in the observational constraints discussed in Section~\ref{sec:analysis}.  To put these constraints into context, let us use the analytic result of this section to relate changes in $\Nf$ to shifts $\delta \ell$ in the peaks of the CMB spectra.  As we show in Appendix~\ref{app:pol}, the E-mode spectrum will exhibit precisely the same phase shift as the temperature spectrum and, therefore, our analytic estimates are applicable in either case. In the flat-sky approximation, a shift in angle $\theta$ is related to a multipole shift by $\delta \ell \simeq (\theta/\pi)\, \Delta\ell_{\text{peak}}$, where $\Delta\ell_{\text{peak}} \sim 330$~\cite{Aghanim:2015wva} is the distance between peaks in the temperature anisotropy spectrum for modes entering the horizon during radiation domination.  Using~\eqref{eq:thetaX}, with $\Nf = N_\nu = 3.046$, we find that the shift of the peaks arising from ordinary neutrinos is $\delta \ell_\nu \approx 26$ (compared to a neutrinoless universe).  Similarly, small variations in $\Nf$ around the standard value will lead to a multipole shift of order
\beq
\delta \ell_{\delta \Nf}  \approx 5.0 \times \delta \Nf \, , \label{eq:delta-ell}
\eeq 
where we have expanded to linear order in $\delta\Nf \equiv \Nf - 3.046$.  While this result is likely subject to a 20 to 30 percent error, it is reliable enough to see that a sensitivity of $\sigma(\Nf) \sim 0.1$ will constrain a phase shift of order $\delta \ell \lesssim 1$.  Current constraints on $\Nf$ imply $\delta \ell \sim {\cal O}(1)$.  As we will see in Section~\ref{sec:analysis}, future CMB experiments are expected to constrain, or measure, shifts of order $\delta \ell \sim {\cal O}(0.1)$. This is consistent with the rough expectation from measuring ${\cal O}(10)$ peaks and troughs in the E-mode power spectrum.

%%%%%%%%%%%%%%%%%%%%%%%%%%%%%
\section{BSM with the CMB: Present and Future}
\label{sec:analysis}
%%%%%%%%%%%%%%%%%%%%%%%%%%%%%

The CMB has the potential to distinguish between many distinct sources of BSM physics: new free-streaming or non-free-streaming particles, isocurvature perturbations and/or non-standard thermal histories.  However, ultimately the observability of the new physics depends both on the size of the effect and whether it is degenerate with other cosmological parameters.  In this section, we present new constraints on the density of free-streaming and non-free-streaming radiation from the Planck satellite~\cite{Aghanim:2015wva} and then discuss the capabilities of a proposed CMB-S4 mission~\cite{Abazajian:2013oma}.  Whenever possible, we will give some approximate analytic understanding of the qualitative origin of our results.  For precise quantitative results, we will perform a full likelihood analysis.\footnote{We prefer the use of MCMC techniques over Fisher matrix forecasts because Fisher matrices can underestimate the impact of degeneracies on the posterior distributions~\cite{Perotto:2006rj}.  We believe this to be the origin of the (small) differences between our results and those of ref.~\cite{Wu:2014hta}.}

\subsection{A Discussion of Degeneracies}
\label{sec:degeneracies}

It is useful to anticipate the possible degeneracies between the effects of extra relativistic species and changes in the cosmological parameters. Here, we show that the degeneracy with the helium fraction $Y_p$ can be understood analytically.

\vskip 4pt
At the level of the homogeneous background, the largest effect of relativistic particles is a change of the expansion rate during radiation domination,
\beq
3 M_{\rm pl}^2 H^2 = \rho_\gamma \left(1+\frac{\Nf + \Nn}{a_\nu}\right)\, ,
\eeq
where free-streaming particles ($\Nf$) and non-free-streaming particles ($\Nn$) contribute equally.  The change in the Hubble rate manifests itself in a modification of the damping tail of the CMB.  Specifically, for modes with wavelengths smaller than the mean free path of the photons, $\lambda \lesssim \lambda_{\rm mfp}$, the temperature fluctuations are suppressed by $\exp[-(k/k_d)^2]$, where $k_d$ is the wave number associated with the mean squared diffusion distance at recombination~\cite{Zaldarriaga:1995gi},
\beq
k_d^{-2} \equiv \int^{a_\rec}_0 \frac{\d a}{a^3 \sigma_T n_e H} \frac{R^2 + \tfrac{16}{15}(1+R)}{6(1+R)^2} \, ,
\label{eq:damping}
\eeq
where $a_\rec$ is the scale factor at recombination, $R$ is the ratio of energy in baryons to photons, $n_e$ is the number density of free electrons and $\sigma_T$ is the Thompson cross section.  Understanding the precise impact of a change in $\Nf+\Nn$ is non-trivial~\cite{Hou:2011ec}, since changing $H$ will also affect the location of the first acoustic peak, which is extremely well measured.  Instead, we must simultaneously vary $H$ and $H_0$, while keeping the angular scale of the first acoustic peak fixed.  Nevertheless, the physical origin of the effect on the damping tail is still given by~(\ref{eq:damping}). This allows us to understand degeneracies with other cosmological parameters.

As pointed out in~\cite{Bashinsky:2003tk}, there is an important degeneracy between $H$ and the helium fraction~$Y_p$.  Since helium has a much larger binding energy than hydrogen, increasing (decreasing) the helium fraction will decrease (increase) the number of free electrons at the time of recombination, $n_e \propto (1-Y_p)$.  It is then possible to change $Y_p$ and $H$ simultaneously in a way that keeps the damping scale fixed, cf.~$k_d^{-2} \propto (n_e H)^{-1}$.  We therefore expect that the CMB temperature constraints on $\Nf+\Nn$ and $Y_p$ to weaken considerably if we allow both of these parameters to vary.   Of course, we also must break the degeneracy between $\Nf$ and $\Nn$, which are not distinguished by their effects on the damping tail.

\vskip 4pt
Fortunately, future data sets will be much less sensitive to these degeneracies for two reasons:  
\begin{itemize}
\item First, as we show in Appendix~\ref{app:pol}, the amplitude of the polarization of the CMB, $\Theta_{P,\ell}$, is proportional to $n_e^{-1}$, but not $H$, and, therefore, it is sensitive to $Y_p$ alone.  The key feature is that polarization is a direct measurement of the quadrupole at the surface of last-scattering, while the damping tail of the temperature spectrum is the integrated effect of the quadrupole on the monopole.  This difference allows us to break the degeneracy between $Y_p$ and $\Nf+\Nn$.

\item Second, as we demonstrated in Section~\ref{sec:analytics}, the CMB is sensitive to the perturbations in the free-streaming particles and not just their contribution to the background evolution.  This is illustrated in Fig.~\ref{fig:CMBphase} and will be explored in more detail in the next subsection.  What is important here is that the phase shift associated with free-streaming particles is not expected to be degenerate with other effects and it is measurable out to very high multipoles.  Furthermore, as we discuss in Appendix~\ref{app:pol}, the same phase shift appears in both temperature and polarization which means there is room for significant improvement in the sensitivity to this effect.  This will largely eliminate the issues of degeneracies for~$\Nf$.
\end{itemize}

\subsection{Detecting Free-Streaming Particles}
\label{ssec:MCMC}

As we move towards more sensitive experiments, the perturbations in the radiation density will play an increasingly important role.  In this section, we will demonstrate the potential of current and future experiments to detect the free-streaming nature of relativistic species.

\subsubsection{Methodology}

The following data were used in our analysis:
\begin{itemize}
\item The Planck likelihoods were described in detail in \cite{Aghanim:2015wva}.  The low-$\ell$ likelihoods ($2\leq \ell \leq 29$) include both the temperature and polarization data (even in the cases labeled ``TT-only"). For the remaining multipoles ($\ell \geq 30$), we use the {\sf Plik} joint TT+TE+EE likelihood.  This contains information about the TT spectrum up to $\ell_{\rm max} = 2508$ and about the TE and EE spectra up to $\ell_{\rm max} = 1996$. The lensing potential is reconstructed in the multipole range $40\leq \ell \leq 400$ using {\sf SMICA} temperature and polarization maps.  For the TT-only constraints, we use the {\sf Plik} TT-only likelihood with range $30\leq \ell \leq 2508$ and the lensing reconstruction involves only the temperature map. 

\item Our forecasts for CMB-S4 experiments assumed $10^6$ polarization-sensitive detectors with a \SI{1}{arcminute} beam and sky coverage $f_{\rm sky} =0.75$.  The default observing time was chosen to be five years, resulting in a sensitivity of $\sigma_T=\sigma_P/\sqrt{2}=0.558 \,\mu {\rm K}\,{\rm arcmin}$ matching the most optimistic experimental setup studied in~\cite{Wu:2014hta}. Our analysis included multipoles up to $\ell_\mathrm{max}=5000$ for both temperature and polarization. We also studied how our results change if we vary the beam size and the maximum multipole. 
\end{itemize}

Our modification of the Boltzmann code {\sf CLASS}~\cite{Blas:2011rf}  includes an additional relativistic fluid, whose energy density is measured by the parameter $\Nn$ defined in (\ref{eq:Neff2}).  The equation of state and the sound speed of the fluid were fixed to $w_Y=c_Y^2= \frac{1}{3}$, with initial conditions that were chosen to be adiabatic.  With this choice, our analytic results imply that this fluid does not contribute to the phase shift in the acoustic peaks.  We use {\sf Monte Python}~\cite{Audren:2012wb} to derive constraints on the parameters $\Nf$ and $\Nn$.  Whenever the primordial helium abundance $Y_p$ was not varied independently (which we will refer to as ``$Y_p$ fixed"), it was set to be consistent with the predictions of BBN, using the total relativistic energy density including both $\Nf$ and $\Nn$ in determining the expansion rate.  All chains were run until the variation in their means was small relative to the standard deviation (using $R-1 \lesssim 0.01$ in the Gelman-Rubin criterion).

\begin{figure}[t!]
\begin{center}
\includegraphics[scale=1.]{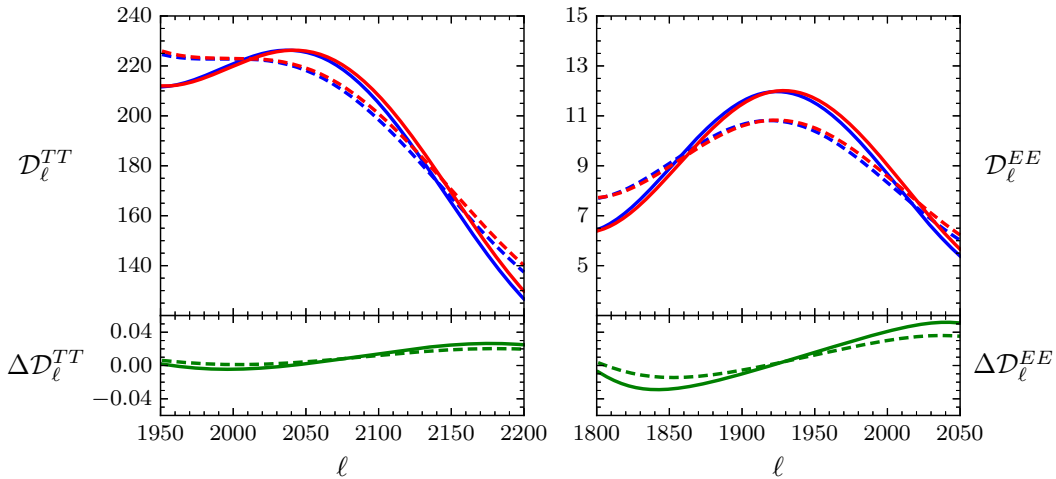}
\caption{{\it Top:} TT spectrum $\mathcal{D}_\ell^{TT}$ ({\it left}) and EE spectrum $\mathcal{D}_\ell^{EE}$ ({\it right}) for ($\Nf=3.046\hskip 1pt, \Nn=0$) (blue) and ($\Nf = 2.046\hskip 1pt, \Nn=1.0$) (red) with $\mathcal{D}_\ell \equiv \ell (\ell+1) C_\ell / (2\pi)$  in units of $\mu{\rm K}^2$. The TT and EE spectra represented by the red curves were rescaled by the same constant factor chosen such that the height of the seventh peak of the TT spectrum matches for the red and blue curves.  The solid and dashed lines show the unlensed and lensed data, respectively.  The phase shift from $\Nf$ and the peak smearing from lensing can be seen in both the TT and EE spectra.
{\it Bottom:} Illustration of the relative difference $\Delta \mathcal{D}_\ell \equiv \delta\mathcal{D}_\ell / \mathcal{D}_\ell$ between the ($\Nf=3.046\hskip 1pt, \Nn=0$) and ($\Nf = 2.046\hskip 1pt, \Nn=1.0$) spectra of the upper panels.  The green solid and dashed lines are the differences in the unlensed and lensed data, respectively.  We see that the change is largest in the unlensed EE spectrum.} 
\label{fig:CMBphase}
\end{center}
\end{figure}

\vskip 4pt
Our analysis makes use of the effects of gravitational lensing of the CMB in two distinct ways.  ``Lensing reconstruction" will refer to a reconstruction of the power spectrum of the lensing potential from the measurements of the temperature and polarization four-point functions. In the case of the CMB-S4 forecasts, the power spectrum of the lensing potential was computed with \textsf{CLASS}.  CMB lensing also modifies the observed CMB power spectra (TT, TE, EE), primarily in the form of smearing the peaks~\cite{Seljak:1995ve}, as is illustrated in Fig.~\ref{fig:CMBphase}.  ``Delensing" removes the effect of lensing on these power spectra using the reconstructed lensing potential.  This is trivially implemented in forecasts in the limit of perfect delensing (we will just output spectra without computing the lensing), but is an involved procedure to implement on real data.  The utility of this procedure is that lensing moves information from the power spectra to higher-point functions, but delensing moves this information back to the power spectra, so that it can easily be accounted for in our likelihood analysis (rather than through some more elaborate multi-point function likelihood).  For a detailed review of gravitational lensing of the CMB we refer to \cite{Lewis:2006fu}.

Our Planck 2015 results use the publicly available lensing reconstruction likelihood, but do not include any delensing of the power spectra.  For CMB-S4, the lensing reconstruction noise was computed using the iterated delensing method described in~\cite{Smith:2010gu} (based on~\cite{Hu:2001kj, Okamoto:2003zw, Hirata:2003ka}).  Forecasts using delensed spectra assumed perfect delensing, which is a good approximation for a CMB-S4 experiment across a wide range of multipoles. Taking lensing reconstruction noise into account when computing delensed spectra would require a more careful analysis. 

\vskip 4pt
The fiducial cosmology used for all forecasts is described by the following parameters: $\Omega_b h^2=0.022$, $\Omega_c h^2=0.120$, $h=0.67$, $A_s=2.42\times10^{-9}$, $n_s=0.965$, $\tau=0.078$, and $\Nf=3.046$.

\subsubsection{Planck 2015 Results}

\begin{figure}[t!]
\begin{center}
\includegraphics[scale=1.]{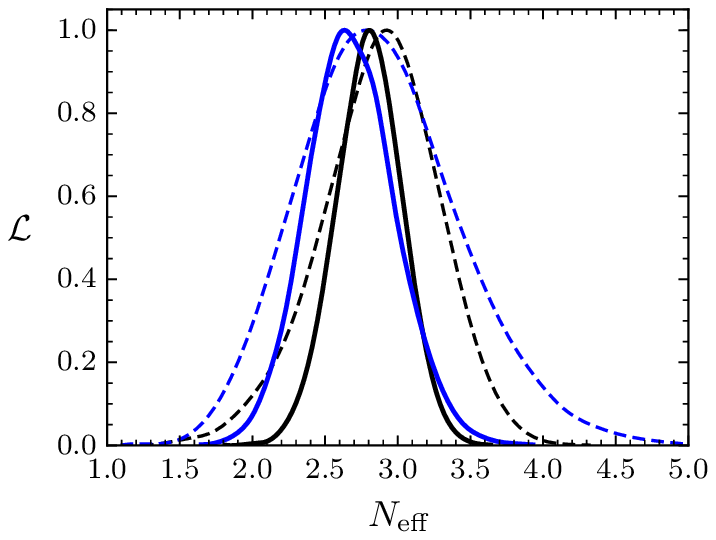} \hskip 15pt
\includegraphics[scale=1.]{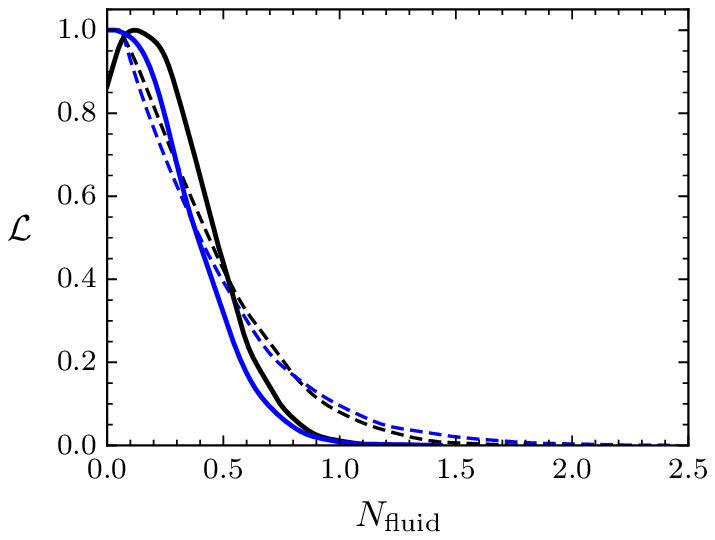}
\vskip -8pt
\caption{{\it Left:} Posterior distributions for $\Nf$ from Planck TT, TE, and EE marginalized over~$\Nn$. The black curve involves the marginalization over $Y_p$, while the blue curve keeps $Y_p$ fixed.  Both likelihoods rule out $\Nf = 0$ at high significance.
{\it Right:} Posterior distributions for $\Nn$ from Planck TT, TE, and EE marginalized over~$\Nf$. The black curve involves the marginalization over $Y_p$, while the blue curve keeps $Y_p$ fixed. 
In both panels, the likelihoods for Planck TT-only with the same marginalizations are shown as dashed lines.}
\label{fig:Nlike15}
\end{center}
\end{figure}

\begin{figure}[t!]
\begin{center}
\includegraphics[scale=1.]{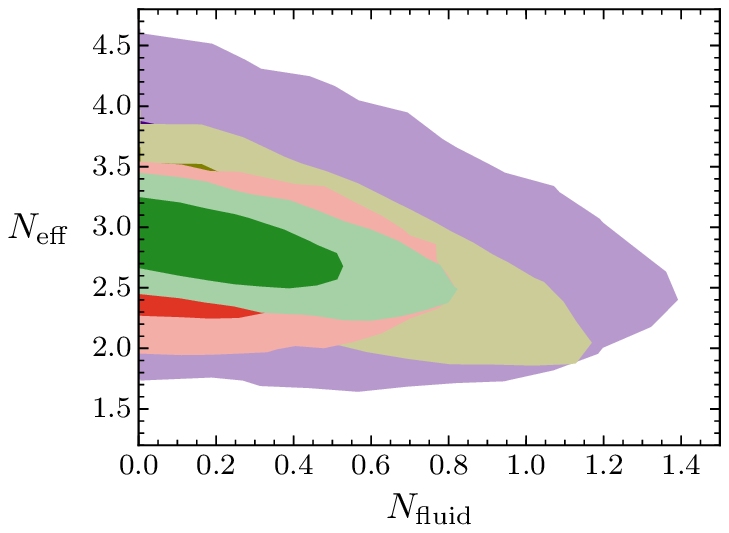} \hskip 15pt
\includegraphics[scale=1.]{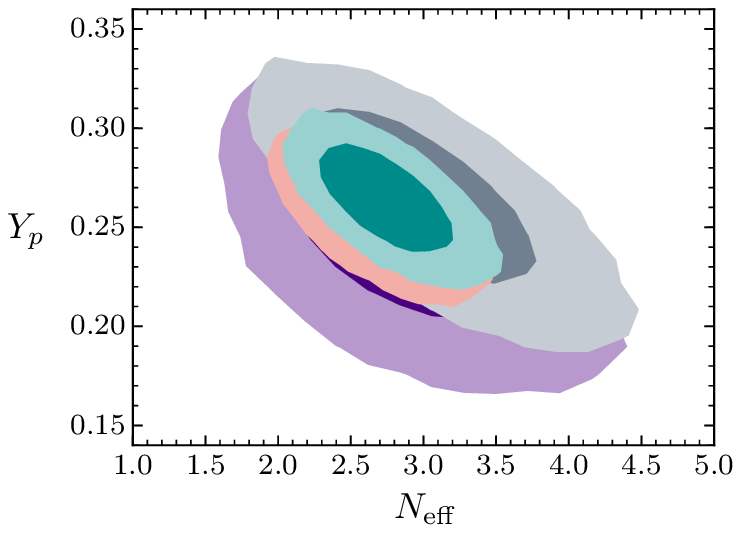}
\caption{{\it Left}: Constraints on $\Nf$ and $\Nn$ using the Planck TT, TE, and EE likelihoods for varying $Y_p$~(red) and when $Y_p$ is fixed~(green).  Shown are also the Planck TT-only results for varying $Y_p$~(indigo) and when $Y_p$ is fixed~(olive).
{\it Right}: Constraints on $\Nf$ and $Y_p$ using the Planck TT, TE, and EE likelihoods for varying $\Nn$~(red) and when $\Nn$ is fixed~(cyan). Shown are also the Planck TT-only results for varying $\Nn$~(indigo) and when $\Nn$ is fixed~(gray).
In both panels, the lensing reconstruction and low-P likelihoods were used for all of the constraints.} 
\label{fig:NeffNfluid}
\end{center}
\end{figure}

The Planck 2015 results have reached an important threshold.  The level of sensitivity is now sufficient to detect the free-streaming nature of the neutrinos (or any additional dark radiation).  In Table~\ref{tab:Planck15}, we present marginalized constraints on $\Nf$ and $\Nn$, with their posterior distributions shown in Fig.~\ref{fig:Nlike15}. We show results both for the combined TT, TE, and EE likelihoods, and for TT alone. The two-dimensional joint constraints are presented in Fig.~\ref{fig:NeffNfluid}. In each case, we compare the results for fixed $Y_p$ with those when $Y_p$ is allowed to vary.  These results robustly demonstrate that there is very little degeneracy between $\Nf$, $\Nn$, and $Y_p$ when using both temperature and polarization data from Planck.
\begin{table}[h!t]
\begin{center}
 \begin{tabular}{c cc cc}
 \toprule
		& \multicolumn{2}{c}{TT, TE, EE}		      				& \multicolumn{2}{c}{TT-only}						\\
 \cmidrule(lr){2-3}\cmidrule(lr){4-5}
	  	& varying $Y_p$ 	    		& fixed $Y_p$ 			& varying $Y_p$			& fixed $Y_p$				\\
 \midrule
   $\Nf$	& $2.68^{+0.29}_{-0.33}$	& $2.80^{+0.24}_{-0.23}$ 	& $2.89^{+0.49}_{-0.62}$	& $2.87^{+0.45}_{-0.37}$	\\ 
   $\Nn$	& $< 0.64$		    		& $< 0.67$  	 			& $< 1.08$	       			& $< 0.94$				\\
 \bottomrule
 \end{tabular}
\caption{Best-fit values and 1$\sigma$ errors for $\Nf$ and 2$\sigma$ upper limits for $\Nn$ for the Planck~2015 data.  Both $\Nn$ and $\Nf$ are allowed to vary in all cases.  The lensing reconstruction and low-P likelihoods were used for all of the constraints.}
\label{tab:Planck15}
\end{center}
\end{table}

From the left panels in Figures~\ref{fig:Nlike15} and~\ref{fig:NeffNfluid} we see that the constraints on $\Nf$ are largely insensitive to the marginalization over $Y_p$ and/or $\Nn$, even when the polarization data is removed.  Since $\Nf$ is degenerate with $Y_p$ and $\Nn$ in the damping tail, this robustness of the constraints suggests that they are driven by the phase shift.  Since $\Nf$ is the unique parameter capable of producing the phase shift, the measurement of the latter breaks the degeneracy between $\Nf$ and both $\Nn$ and $Y_p$.  We also see that adding polarization data leads to a large improvement in the constraint on $\Nf$, most likely because the peaks of the E-mode spectrum are sharper, which makes the phase shift easier to measure~\cite{Bashinsky:2003tk}.  This is illustrated in Fig.~\ref{fig:CMBphase}, where we show the relative differences in the TT and EE spectra when varying $\Nf$ and $\Nn$.  While the phase shift is visible in both cases, the size of the effect is larger in the polarization spectrum which increases the impact of the E-mode data.  

Similar analyses were performed in~\cite{Bell:2005dr,Friedland:2007vv} using WMAP data (and external datasets).  Their results are qualitatively similar to our TT-only analysis with $Y_p$ fixed, although with weaker constraints on $\Nf$ and $\Nn$.  By comparison, adding E-mode data further improves constraints in the $\Nf$-$\Nn$ plane, also when $Y_p$ is allowed to vary.

Figure~\ref{fig:NeffNfluid} shows that there is a significant difference in the constraints in the $\Nf$-$Y_p$ plane, with and without the polarization data.  Without polarization, allowing $\Nn$ to vary weakens constraints on $\Nf$ and $Y_p$.  This is consistent with our discussion of degeneracies in \textsection\ref{sec:degeneracies}.  Using only the temperature data, $Y_p$ and $\Nn$ are measured mostly from the damping tail, but their effects on the latter are degenerate.  As a result, the constraints on $Y_p$ and $\Nf$ weaken when $\Nn$ is allowed to vary.  The situation changes when polarization data is added. Now, there is very little difference in the constraints as we vary the marginalization over additional parameters.  This is most noticeable in the right panel in Fig.~\ref{fig:NeffNfluid}, where the constraints on $Y_p$ and $\Nf$ become nearly independent of the treatment of $\Nn$.  This feature was anticipated in \textsection\ref{sec:degeneracies}, where it was observed that polarization breaks the degeneracy between $Y_p$ and $\Nn$.

\vskip 4pt
In~\cite{Follin:2015hya}, a constraint was recently placed on the effective number of free-streaming species by isolating the phase shift in the Planck 2013 temperature data: $\Nf = 2.3^{+1.1}_{-0.4}$ ($68\%$\hskip 2ptC.L.) while keeping the damping tail fixed and $\Nf = 3.5\pm0.65$ ($68\%$\hskip 2ptC.L.) when marginalizing over the effect on the damping tail~\cite{FollinPC}.  To compare to that analysis, we remove the polarization data. We then find $\Nf =2.89^{+0.49}_{-0.62}$\hskip 1pt, which is quite similar to the direct measurement of the phase shift.\footnote{Due to the marginalization over $\Nn$ and $Y_p$, our analysis is most comparable to the marginalized result: $\Nf = 3.5\pm0.65$~\cite{FollinPC}. One difference in our approach is that it includes information in the amplitude shift produced by the free-streaming species. However, this effect is likely subdominant to the phase shift due to the degeneracy with the amplitude of the primordial power spectrum.  As we will discuss in the next subsection, when we allow $Y_p$ to vary, future CMB missions get the majority of the sensitivity of $\Nf$ from the phase shift and so we expect our methods to produce increasingly similar results.}  When we add the TE and EE likelihoods, our constraint improves by about a factor of two to $\Nf = 2.68^{+0.29}_{-0.33}$.  From the estimate\hskip 1pt\footnote{We can also estimate this more schematically from the knowledge that $\Nf=3.046$ produces $\delta \ell \approx 10$ for $\ell \lesssim 3000$ relative to $\Nf = 1$~\cite{Follin:2015hya}. Current constraints allow for roughly a 10 percent variation in $\Nf$, which would imply $\delta \ell \approx 1$.  Direct measurements of the individual peak locations are given in~\cite{Aghanim:2015wva} with a similar level of precision.} in (\ref{eq:delta-ell}), we conclude that these constraints correspond to a phase shift of about $\delta \ell \approx 1$.  While this is compatible with expectations,\footnote{Forecasts using the isolated phase shift alone give $\sigma(\Nf) = 0.41$ for Planck with polarization~\cite{Follin:2015hya}.} it is nonetheless impressive that the data is sensitive to these small and subtle effects.

\vskip 4pt
We note in passing that our results show that $Y_p$ is somewhat higher than its expected value.  This is evident in Fig.~\ref{fig:NeffNfluid}, where the value of $Y_p$ is almost 1$\sigma$ higher than the BBN prediction of $Y_p \approx 0.247$. This was also observed by the Planck collaboration~\cite{Ade:2015xua}, although their values of $Y_p$ are somewhat lower.\footnote{We found results closer to the Planck values when using {\sf CosmoMC}/{\sf CAMB} \cite{Lewis:2002ah, Lewis:1999bs} rather than {\sf Monte Python}/{\sf CLASS} in the analysis with varying $\Nf$ and $Y_p$.  There are several subtle differences in the implementations of these codes which could be responsible for this disagreement.}  It remains unclear what precisely in the data is driving $Y_p$ to larger values.  This should serve as a reminder that as we increase precision, we will only be able to trust a BSM interpretation of our data if accuracy is addressed at the same level.

\subsubsection{CMB Stage IV Forecasts}

While current data is already sensitive to the free-streaming nature of neutrinos, future experiments are expected to improve these constraints by at least an order of magnitude.  As we have emphasized in the introduction, increasing the sensitivity to the $\sigma(\Nf) \sim 10^{-2}$ level probes a number of plausible BSM scenarios that are currently unconstrained.  Preliminary forecasts~\cite{Wu:2014hta} suggest that this level is indeed achievable.  Our goal is to extend these results in two ways: (\hskip -0.5pt{\it i}\hskip 1pt) to include $Y_p$ and $\Nn$ to identify more clearly the types of BSM physics we might be sensitive to and (\hskip -0.5pt{\it ii}\hskip 1pt) to perform a full likelihood analysis (rather than a Fisher forecast) to ensure that degeneracies are treated correctly (see~\cite{Perotto:2006rj} for a discussion).

\begin{table}[t!]
\begin{center}
 \begin{tabular}{l cc ccc} 
 \toprule
   Experiment  & Delensing	& Reconstruction	& $\sigma(\Nf)$ 			& $\sigma(Y_p)$     		& $\Nn$			\\[0.5ex] 
 \midrule
   Planck 2015 & No 		& Yes			& 0.31\phantom{0}	& 0.019\phantom{0} 	& $< 0.64$	\\ 
 \cmidrule{1-6}
	       		& No 		& No			& 0.062 	     				& 0.0053					& $< 0.18$		\\ 
   CMB-S4      	& Yes		& No			& 0.054 	     				& 0.0044   				& $< 0.17$ 		\\
	       		& Yes		& Yes			& 0.050 	     				& 0.0043   				& $< 0.16$ 		\\
 \bottomrule
\end{tabular}
\caption{Marginalized likelihoods for Planck and CMB-S4.  All likelihoods allow $\Nf$, $\Nn$, and~$Y_p$ to vary.  The Planck likelihood uses the TT, TE, and EE power spectra, as well as the lensing reconstruction likelihood.  The forecasts for CMB-S4 include the possibilities of delensing and lensing reconstruction. We excluded the forecast using lensed spectra combined with lensing reconstruction since this is known to produce overly optimistic error forecasts due to a double counting of lensing information~\cite{Hu:2001fb}.  The double counting can be safely ignored for Planck, but will become more important for future experiments~\cite{Schmittfull:2013uea}.  Displayed are $1\sigma$ error bars for $\Nf$ and $Y_p$, and $2\sigma$ upper limits for $\Nn$.}
\label{tab:forecasts}
\end{center}
\end{table}

\begin{figure}[t!]
\begin{center}
\includegraphics[scale=1.]{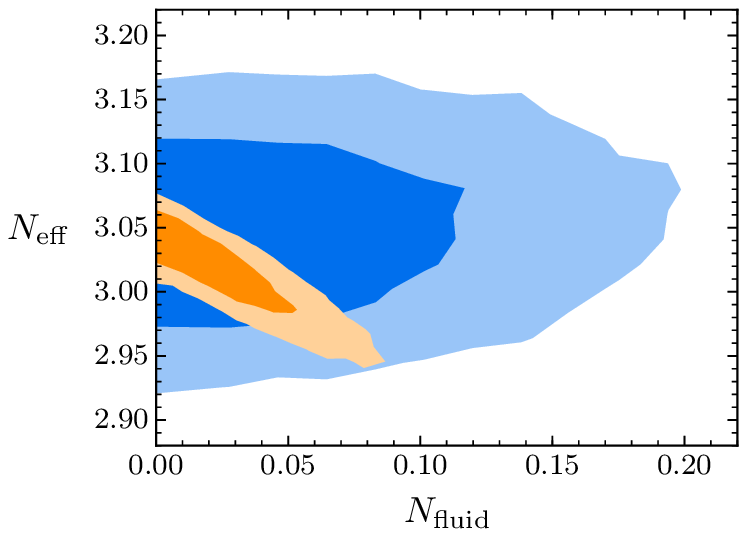}\hskip 15pt
\includegraphics[scale=1.]{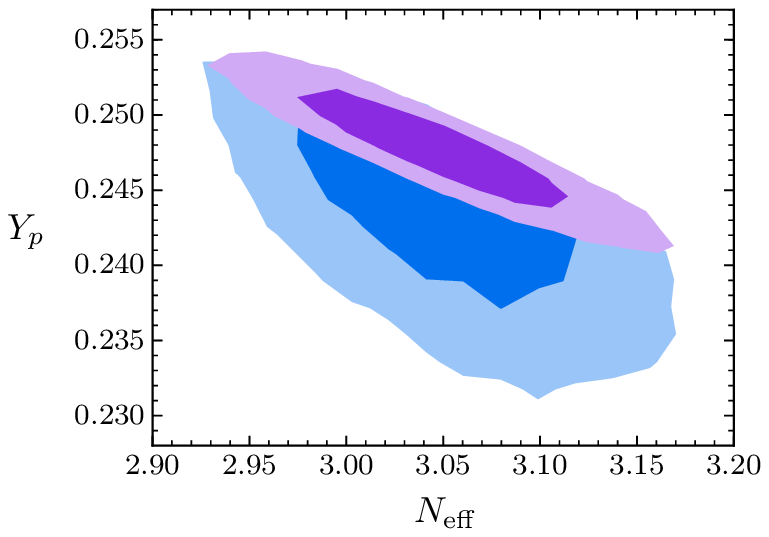}
\vskip -8pt
\caption{{\it Left}: Forecasted CMB-S4 constraints on $\Nf$ and $\Nn$ for $Y_p$ fixed (orange) and varying~(blue).
{\it Right}: Forecasted CMB-S4 constraints on $\Nf$ and $Y_p$ for $\Nn = 0$ fixed (purple) and varying~(blue).} 
\label{fig:NnNf}
\end{center}
\end{figure}

\vskip 4pt
The results of our forecasts are summarized in Table~\ref{tab:forecasts}.  Given the constraints from Planck, it is not surprising that $\Nf$ is easily distinguished from $\Nn$ with CMB-S4 experiments.  As before, the constraints on $\Nf$ are significantly stronger than those on $\Nn$, which is consistent with the interpretation that these parameters are being distinguished by differences in the perturbations for the two types of radiation.  When both radiation components are included, the detailed matching of the acoustic peaks is very sensitive to the phase shifts due to $\Nf$, but is much less affected by $\Nn$. This is illustrated by the left panel in Fig.~\ref{fig:NnNf}, which presents the joint constraints on $\Nn$ and $\Nf$, both for fixed $Y_p$ and when it is allowed to vary.  When $Y_p$ is fixed, there is a strong degeneracy between $\Nf$ and $\Nn$ which is absent when $Y_p$ is allowed to vary.  Yet, in both cases, the contours close at roughly the same value of $\Nf$, indicating that the degeneracy is broken in a way that is insensitive to the damping tail.

As we have emphasized throughout, the strong constraint on $\Nf$ arises in part from the sharpness of the peaks of the E-mode spectrum, which leads to better measurements of the phase shift.  This same intuition explains why the constraints on $\Nf$ are strengthened by using the delensed power spectrum.  One of the well-known effects of lensing is a smearing of the acoustic peaks~\cite{Seljak:1995ve} (see Fig.~\ref{fig:CMBphase}). By delensing the power spectra, we sharpen the peaks, which makes the measurements of the phase shift more precise.  We illustrate the impact that delensing can have in Fig.~\ref{fig:CMBphase}, where we see that the effect of changing $\Nf$ produces a much larger relative change on the unlensed data.  This procedure is analogous to reconstructing the BAO peak to sharpen distance measurements~\cite{Eisenstein:2006nk}.  In both cases, non-linearities transfer information from the power spectrum to higher-point correlation functions.  Delensing or BAO reconstruction moves that information back to the power spectrum, so that it is more easily accounted for in these analyses.  As a result, the error in the measurement of any quantity that is sensitive to the sharpness of the peak will be reduced (like the phase shift and the BAO scale).  This suggests that delensing will be a useful tool for improving constraints on cosmological parameters in these future experiments. 

\vskip 4pt
While much of our efforts have been devoted to understanding the degeneracies between $\Nf$, $\Nn$, and $Y_p$, the actual physical models we wish to constrain may not exhibit these degeneracies.  First of all, many models with additional light fields still have $\Nn = 0$ (i.e.~only free-streaming radiation)~\cite{Brust:2013ova}, which would in principle allow us to combine information from the damping tail and the phase shift to constrain $\Nf$.  Furthermore, while $Y_p$ is often affected by BSM physics at the time of BBN, the precise degeneracy needed to keep the damping tail fixed is unlikely to occur naturally.  The constraints on such models would be significantly stronger.  As we see in Table~\ref{tab:forecasts2}, a factor of 3 to 4 improvement in the constraints is possible when this degeneracy with changes in $Y_p$ is not present.  
\begin{table}[h!t]
\begin{center}
 \begin{tabular}{l cc} 
 \toprule
   Experiment		& $\sigma(\Nf)$	& $\sigma(Y_p)$     \\ [0.5ex] 
 \midrule
   \multirow{2}{*}{Planck 2015} 		& 0.30\phantom{0} & 0.018\phantom{0} \\
				& 0.19\phantom{0}& --    	     		\\
 \cmidrule{1-3}
   \multirow{2}{*}{CMB-S4}			& 0.048  	  & 0.0027\\
				& 0.013			& --    	   		\\
 \bottomrule
\end{tabular}
\caption{Results for the marginalized $1\sigma$ errors for Planck (TT, TE, EE, lensing reconstruction) and forecasts for CMB-S4 for $\Nf$ and $Y_p$, when $\Nn = 0$ is held fixed.  The CMB-S4 forecasts assumed both delensing and lensing reconstruction.  A dash in the $\sigma(Y_p)$ entry indicates that $Y_p$ was fixed by consistency with BBN. }
\label{tab:forecasts2}
\end{center}
\end{table}

\vskip 4pt
Finally, perhaps the most important feature of these forecasts is the dramatic improvement, relative to Planck 2015, that can be expected from a plausible experimental configuration of CMB-S4.  Our projections suggest that a factor of 5 to 10 improvements are achievable, but we should also investigate the robustness of this conclusion to changes of the experimental configuration.  Here, we study variations of the beam size ($\theta_b$) and the maximal available multipole ($\ell_{\rm max}$).  These are important for two reasons: (\hskip -0.5pt{\it i}\hskip 1pt) the beam size is ultimately a choice made within the context of limited resources and (\hskip -0.5pt{\it ii}\hskip 1pt) the presence of foregrounds or systematics make it difficult to predict $\ell_{\rm max}$ reliably beforehand.  In Table~\ref{tab:beam} we show the forecasts for various values of $\theta_b$ and $\ell_{\rm max}$, assuming $10^6$ detectors. 
\begin{table}[t!]
\begin{center}
\begin{tabular}{l ccc cc} 
 \toprule
   Parameter 				    			& $1'$  		& $2'$  		& $3'$  		& $\ell_{\rm max} = 3000$	& $\ell_{\rm max} = 4000$ \\ [0.5ex]
 \midrule
   $\sigma(\Nf)$ ($Y_p$ fixed, $\Nn =0$)	& 0.013 		& 0.015 		& 0.016		& 0.023 	  	  	& 0.015 		  \\
   $\sigma(\Nf)$ ($Y_p$ fixed, $\Nn \ne 0$)	& 0.026 		& 0.027 		& 0.029		& 0.034 	  	  	& 0.028 		  \\
   $\sigma(\Nf)$ ($Y_p$ varying, $\Nn =0$)   & 0.048 		& 0.051 		& 0.055		& 0.058 	  	  	& 0.052 		  \\
   $\sigma(\Nf)$ ($Y_p$ varying, $\Nn \ne 0$) & 0.050 		& 0.052 		& 0.055		& 0.061 	  	  	& 0.051 		  \\
 \cmidrule{1-6}
   $\Nn$ ($Y_p$ varying)		    & $<0.16\phantom{0}$   	& $<0.17\phantom{0}$ & $<0.18\phantom{0}$ & $<0.20\phantom{0}$ 	& $<0.17\phantom{0}$      \\
   $\Nn$ ($Y_p$ fixed)  		    & $<0.068$	& $<0.072$	& $<0.076$	& $<0.090$	& $<0.072$ 		 			 \\
 \bottomrule
\end{tabular}
\caption{Forecasts for a CMB-S4 experiment with varying beam size and maximum multipole assuming $10^6$ detectors.  We take $\ell_{\rm max} =5000$ as we vary the beam size and $\theta_b =1'$  when we vary $\ell_{\rm max}$.  Displayed are $1\sigma$ error bars for $\Nf$ and $2\sigma$ upper limits for $\Nn$.}
\label{tab:beam}
\end{center}
\end{table}

There are two simple lessons we can draw from this table.  First of all, measuring modes with $\ell > 3000$ offers only a moderate improvement on our constraints.  Similarly, the benefit to reducing the beam size significantly is also limited.  Most of the improvement in the measurement of the phase shift is coming from high-precision measurements of E-modes with $\ell < 3000$.  In contrast, when $Y_p$ is fixed and $\Nn=0$, the sensitivity to the beam size and $\ell_{\rm max}$ is stronger, which suggests\hskip 1pt\footnote{We have not been careful to account for foregrounds in temperature or polarization.  Information in our forecasts coming from high-$\ell$ temperature data is unlikely to be available in a real experiment.} that we are gaining useful information from the damping tail at $\ell > 3000$.

We have fixed the number of detectors to be $10^6$, but the precise value will play a very important role in the ultimate reach of CMB-S4, not just for $\Nf$, but for most physics targets.  For $\Nf$ specifically, we are still far from the limit set by cosmic variance in the E-mode power spectrum for $\ell \gtrsim 1000$, which is where most of the improvements in the phase shift measurements are coming from.  As a result, within the range of $10^4$ to $10^6$ detectors being considered, we improve constraints significantly by further reducing the detector noise and, hence, increasing the number of detectors.  Forecasts for $\Nf$ with varying numbers of detectors were studied in~\cite{Wu:2014hta}, which confirm this intuition.

\subsection{Time Evolution of Radiation Densities}
\label{sec:timeEvol}

One of the important benefits of measuring the fluctuations in the dark radiation is that it eliminates the degeneracy between $Y_p$ and $\Nf$.  This has important consequences for constraints on BSM physics, since $Y_p$ is sensitive to the total radiation density at the time of BBN (\SI{3}{\minute} after the Big Bang), while $\Nf$ and $\Nn$ are related to these radiation densities around the time of recombination (\num{380000} years after the Big Bang).  As a result, CMB measurements of $Y_p$ and $\Nf$ probe scenarios where these densities change between those two times, perhaps through the decay of a heavy particle or through some other production mechanism~\cite{Fischler:2010xz}.

\vskip 4pt
It is useful to translate constraints on $Y_p$ into constraints on the radiation density at the time of BBN~\cite{Bernstein:1988ad},
\beq
Y_p \approx 0.247 + 0.014 \times \delta N_{\rm eff+fluid}^{\rm BBN}\, , \label{eq:YpBBN}
\eeq
where $\delta N_{\rm eff+fluid}^{\rm BBN} \equiv \Nf^{\rm BBN}+\Nn^{\rm BBN} - 3.046$.  Marginalizing over $\Nn$, we found an error in the helium fraction of $\sigma(Y_p) =0.0043$ for CMB-S4 (see Table~\ref{tab:forecasts}), while for fixed $\Nn=0$,  we get $\sigma(Y_p) = 0.0027$ (cf.\ Table~\ref{tab:forecasts2}).  Using~(\ref{eq:YpBBN}), these constraints imply
\beq
\sigma(N_{\rm eff+fluid}^{\rm BBN}) = \left\{\begin{array}{ll} \ 0.31 & \quad \Nn \ne 0 \\[6pt] \ 0.19 & \quad \Nn = 0 \end{array}  \right.\, .
\eeq
The last constraint is stronger than the current best limit of $\sigma(N_{\rm eff+fluid}^{\rm BBN}) = 0.28$ from BBN alone~\cite{Cyburt:2015mya}. The CMB will therefore provide independent measurements of $\Nf$ at two different times in a single experiment, each surpassing our current level of sensitivities from combining multiple probes.

\vskip 4pt
While the constraint on $N_{\rm eff+fluid}^{\rm BBN}$ from a CMB-S4 experiment is only a modest improvement over current measurements from primordial abundances, it has the unique advantage that it is a clean measurement (i.e.~it is not affected by astrophysical processes at later times) and it can be combined with measurements of other cosmological parameters (e.g.~$\Omega_b h^2$) without combining different data sets.  A common approach with current data is to combine the constraints from the CMB and primordial abundances in order to improve the overall sensitivity to $\Nf$, in the case where it is time independent.  From Table~\ref{tab:forecasts2}, we see that if we do not allow a variation in $\Nf$ between BBN and the CMB, we get very strong constraints on $\Nf$ due to the lack of degeneracies in the damping tail.  These results are sufficiently strong so that it is unlikely that including information from primordial abundances will lead to much improvement.

When discussing the time variation of $\Nf$, we have focused only on the model-independent measurement implied by varying $\Nf$ and $Y_p$ independently.  As a result, the constraints we derive are controlled primarily by the degeneracy between $\Nf$ and $Y_p$ in the damping tail.  Without this degeneracy, the constraints on $\Nf$ are much stronger.  In realistic models, it may be the case that both $\Nf$ and $Y_p$ are changed independently, but that they do not produce this degeneracy in the relevant range of parameters.  For such models, a dedicated analysis of CMB data would likely offer a much larger gain over the current limits from primordial abundances.

\subsection{Implications for BSM Physics} 

In this section, we have focused on simple descriptions of BSM physics in terms of the effective parameters $\Nf$, $\Nn$, and $Y_p$.  These parameters capture important aspects of CMB physics and our cosmological history, which may be used to test a number of scenarios for BSM physics (see Table~\ref{tab:bsmsummary}):
\begin{table}[t!]
\begin{center}
 \begin{tabular}{l ll} 
 \toprule
   \textit{Signature}	& \textit{Influenced by}												& \textit{Degeneracies broken by}	\\
 \midrule\addlinespace
   CMB damping tail	& $\Nf^{\rm CMB}$+$\Nn^{\rm CMB}$\hskip 1pt, $Y_p$\hskip 1pt, $E_{\rm inj}^{\rm CMB}$		& Phase shift, Polarization	\\[4pt]
   Phase shift   	& $\Nf$\hskip 1pt, $N_{\rm fluid}^{\rm iso}$\hskip 1pt, $\cancel{\rm GR}$		& Scale dependence 		\\[4pt]
   Spectral distortions  & $E_{\rm inj}^{\text{post-BBN}}$									&		 				\\[4pt]
   Primordial abundances & $\Nf^{\text{BBN}}+\Nn^{\rm BBN}$\hskip 1pt, $N_\nu$\hskip 1pt, $\eta^{\rm BBN}$\hskip 1pt, $E_{\rm inj}^{\rm BBN}$ & CMB	\\
 \addlinespace[4pt]\bottomrule
\end{tabular}
\caption{Cosmological probes of BSM physics and their sensitivity to free-streaming and non-free-streaming radiation ($\Nf$ and $\Nn$), the number of active neutrinos ($N_\nu$), the baryon-to-photon ratio~($\eta$), and the amount of energy injection ($E_{\rm inj}$). The superscripts BBN, CMB, or post-BBN denote the time at which a quantity is being probed, where post-BBN refers to redshifts of $z \lesssim 10^6$ when spectral distortions become possible.  The parameter $\Nn^{\rm iso}$ abstractly stands for isocurvature fluctuations, while $\cancel{\rm GR}$ denotes modified gravity.\vspace{-10pt}}
\label{tab:bsmsummary}
\end{center}
\end{table}

\begin{itemize}\itemsep0pt
\item For minimal extensions of the Standard Model with a light field, e.g.~\cite{Brust:2013ova, Buen-Abad:2015ova, Chacko:2015noa}, $\Nf$ is sensitive to the freeze-out of the particle.  At current levels of sensitivity, $\sigma(\Nf) \gtrsim 0.1$, we can rule out some scenarios where particles freeze-out after the QCD phase transition~\cite{Brust:2013ova}.  Freeze-out before the QCD phase transition typically dilutes the contribution to $\Nf$ by a factor of 10, which allows such models to easily evade current constraints.  Fortunately, these scenarios are likely to be accessible with CMB-S4 experiments~\cite{Chacko:2015noa}.  For these cases, we are sensitive to sufficiently early times so that BSM physics above the \si{\tera\electronvolt} scale may be important. 

\item Measurements of the effective number of free-streaming particles at recombination, $\Nf^{\rm CMB}$, are also sensitive to energy injected into the Standard Model particles after the time of neutrino decoupling ($\sim \SI{0.1}{sec}$).  Depending on the time and nature of this energy injection, it may alter the primordial abundances or introduce spectral distortions which would distinguish it from a new light field.  For example, a decay to photons after BBN would lower $\Nf^{\rm CMB}$ and $\eta^{\rm BBN}$ (the baryon-to-photon ratio at BBN), while keeping the radiation density at BBN, $\Nf^{\rm BBN}$, fixed~\cite{Cadamuro:2010cz}.  

\item Energy injection of many kinds is a typical byproduct of changing $\Nf$, but may also be the dominant signature of BSM physics.   Decays during BBN can disrupt the formation of nuclei without substantially changing the total energy in radiation.  Alternatively, recombination is very sensitive to energy injection~\cite{Padmanabhan:2005es} which can alter the form of the visibility function.\footnote{The common element of both of these examples is that the tail of the Boltzmann distribution is playing a critical role (due to the large value of $\eta^{-1}$).  As a result, the change to the small number of high-energy photons is more important than the total energy density.}  

\item As we discussed in Section~\ref{sec:analytics}, phase shifts of the acoustic peaks may also be produced by isocurvature perturbations (denoted by $\Nn^{\rm iso}$ in the table).  We offered a simple curvaton-like example of this effect, but we expect to be broadly sensitive to physics in the dark sector that is not purely adiabatic.  Since there are many good reasons to imagine why isocurvature perturbations might arise in the dark sector, this motivates a future exploration of the observability of these effects. 

\item Finally, we have assumed the validity of the Einstein equations throughout.  This enforced that $\Phi_- = 0$ in the absence of anisotropic stress. Modified theories of gravity are often parameterized in terms of their change to the Einstein constraint equation and the corresponding effect on $\Phi_-$; see e.g.~\cite{Zhang:2007nk}. Since the field $\Phi_-$ played an important role in our analysis of the phase shift, it would be interesting to explore how the result changes for specific modifications of GR.  Conversely, the phase shift of the CMB spectrum may be an interesting probe of modified gravity.
\end{itemize}

%%%%%%%%%%%%%%%%%%%%%%
\section{Conclusions and Outlook}
\label{sec:conclusions}
%%%%%%%%%%%%%%%%%%%%%%

CMB observations have become precise enough to probe the {gravitational} imprints of BSM physics on the perturbations of the primordial plasma.  In the upcoming era of CMB polarization experiments, our sensitivity to these subtle effects will increase significantly and will offer new opportunities in the search for new physics.  It is therefore timely to re-evaluate how CMB data can inform our view of the laws of physics.

\vskip 4pt
In this paper, we have explored how the phase shift of the acoustic peaks might be used as such a probe.  This phase shift is particularly interesting because analytic properties of the Green's function of the gravitational potential strongly limit the possible origins of such a shift to
\begin{center}
\begin{tabular}{r l c l}
{\it i.} & waves propagating faster than the sound speed of the photon-baryon fluid, \\[4pt]
{\it ii.} & isocurvature fluctuations.
\end{tabular}
\end{center}
\noindent
For adiabatic initial conditions, the phase shift is most easily generated by free-streaming radiation and becomes an excellent measure of the effective number of neutrinos $\Nf$ at the time of recombination.  Realistic models of isocurvature fluctuations typically produce a scale-dependent phase shift, which allows them to be distinguished from changes to the energy density of the radiation.  

\begin{figure}[h!t]
\begin{center}
\includegraphics[scale=1.]{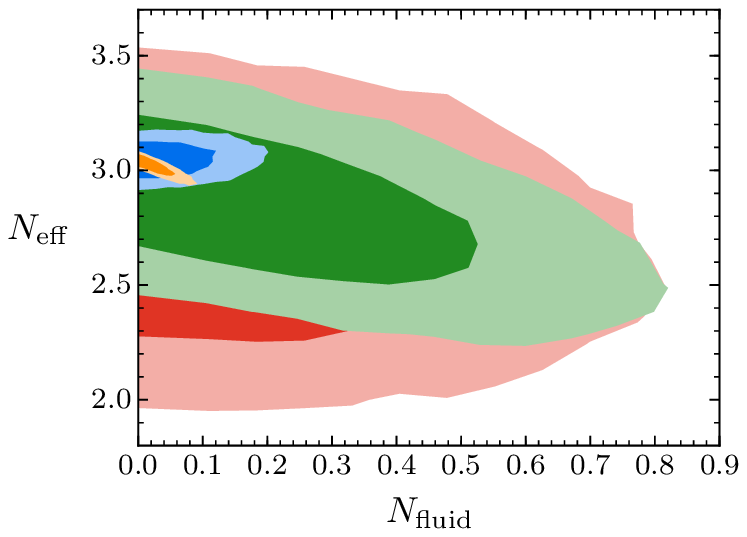} \hskip 10pt
\includegraphics[scale=1.]{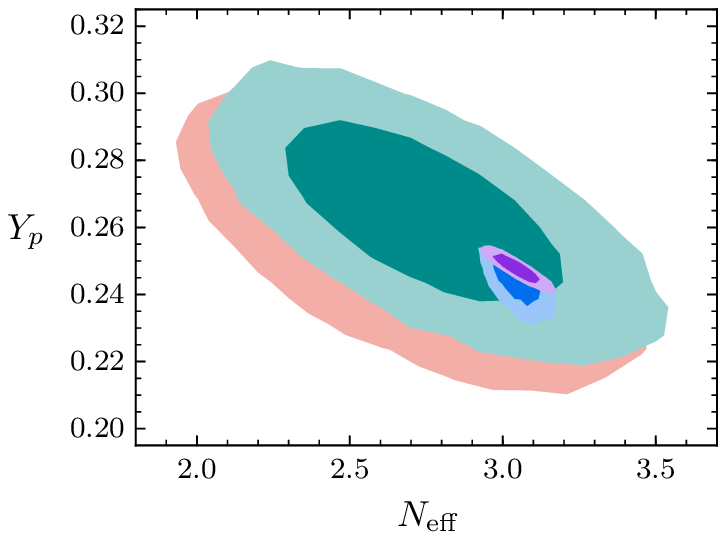}
\vskip -10pt
\caption{{\it Left:} Planck constraints on the effective number of free-streaming and non-free-streaming relativistic species, $\Nf$ and $\Nn$, allowing the helium fraction $Y_p$ to vary~(red) and keeping it fixed~(green).
{\it Right:} Planck constraints on $Y_p$ and $\Nf$ for varying $\Nn$~(red) and when $\Nn=0$ is fixed~(cyan).
In both plots, the contours from the CMB-S4 forecasts, presented in Fig.~\ref{fig:NnNf}, have been included to show the expected improvements in sensitivity.\vspace{-13pt}}
\label{fig:Summary}
\end{center}
\end{figure}

\vskip 4pt
What makes these results particularly compelling is that current and future CMB experiments are sensitive enough to detect these phase shifts at high significance~\cite{Follin:2015hya}.  We have demonstrated this with an analysis of the 2015 data from the Planck satellite and forecasts for a CMB Stage~IV experiment (see Fig.~\ref{fig:Summary}).  Our results include a clear detection of the free-streaming nature of neutrinos.  They also highlight the important role played by the polarization data in breaking the degeneracy between the contributions from free-streaming and non-free-streaming species, $\Nf$ and $\Nn$, as well as that with the helium fraction $Y_p$.

\vskip 4pt
The increased sensitivity to differences in the evolution of perturbations motivates revisiting the predictions of specific models of BSM physics. Reference~\cite{Brust:2013ova} performed a comprehensive study of `minimal' effective field theory realizations of technically natural forms of dark radiation.  However, the predictions of these models were only compared to existing constraints on $\Nf$, i.e.~only modifications to the background density were taken into account.  (Recently, these estimates were updated~\cite{Chacko:2015noa} in light of the higher sensitivity expected in CMB-S4 experiments.)  In the future, we plan to revisit this approach, including the effects of the perturbations.  These effects may arise from new degrees of freedom (e.g.~\cite{Cadamuro:2010cz, Menestrina:2011mz, Boehm:2012gr, Brust:2013ova, Weinberg:2013kea, Cyr-Racine:2013fsa, Vogel:2013raa, Millea:2015qra, Chacko:2015noa}) or from non-standard properties of the SM neutrinos (e.g.~\cite{Cyr-Racine:2013jua, Archidiacono:2013dua, Oldengott:2014qra}). We also hope to explore `non-minimal' extensions of the effective theories of~\cite{Brust:2013ova}.  For example, decays of massive fields are abundant in well-motivated extensions of the Standard Model, but were forbidden in~\cite{Brust:2013ova} on the basis of being non-minimal.  As we illustrated, the type of energy injection that would be produced from such a decay can be observable over a wide range of times, even in the absence of new light fields.  Furthermore, the detailed predictions for $\Nf$ may be significantly altered from the minimal case, which can change the impact of future constraints.

\vskip 4pt
The expected improvement in the constraints on extra relativistic species in future experiments is quite remarkable, e.g.~$\sigma(\Nf) \sim 0.01$. This provides an opportunity to probe BSM physics at a much more precise level than was previously possible.  We are optimistic that this will teach us something interesting.  We will either discover a whole new world of dark physics, or learn to what remarkable degree it is decoupled from the rest of physics.

\vspace{0.21cm}
\subsubsection*{Acknowledgements}

We thank Dick Bond, Nathaniel Craig, Chandrima Ganguly, Lloyd Knox, Daan Meerburg, Zhen Pan and Alex Van Engelen for discussions.  We also thank Kris Sigurdson for bringing ref.~\cite{Bell:2005dr} to our attention and Follin et al.~\cite{FollinPC} for providing the unpublished results discussed in the main text.  We are grateful to Chris Brust, Yanou Cui, and Kris Sigurdson for making us aware of a problem related to our treatment of nuisance parameters for the Planck 2015 likelihoods, which led to weaker Planck constraints in an earlier version of this work. D.\,B.~and B.\,W.~acknowledge support from a Starting Grant of the European Research Council (ERC STG Grant 279617). B.\,W.~is also supported by a Cambridge European Scholarship of the Cambridge Trust and an STFC Studentship. D.\,G.~is supported by an NSERC Discovery Grant.  This work is partly based on observations obtained by the Planck satellite (\href{http://www.esa.int/Planck}{http:/\!/www.esa.int/Planck}), an ESA science mission with instruments and contributions directly funded by ESA Member States, NASA, and Canada.  Parts of this work were undertaken on the COSMOS Shared Memory System at DAMTP (University of Cambridge), operated on behalf of the STFC DiRAC HPC Facility. This equipment is funded by BIS National E-Infrastructure Capital Grant ST/J005673/1 and STFC Grants ST/H008586/1, ST/K00333X/1.

\clearpage
\appendix

%%%%%%%%%%%%%%%%%%
\section{Comments on Matter}
\label{app:ext}
%%%%%%%%%%%%%%%%%%

In the main text, we computed the phase shift of the photon density fluctuations assuming a radiation-dominated background.  While this simplification made an analytic treatment possible, we may wonder if it misses important effects.  In this appendix, we will bridge this gap to the degree which is possible without using numerics, focusing on the contributions from free-streaming radiation.

\vskip 4pt
There are several reasons why we want to understand the contributions to the phase shift from modes in the matter era. First, recombination occurs during matter domination and, therefore, in principle, it could be important for every mode in the CMB. Second, modes corresponding to large angular scales (small $\ell$) enter the horizon during (or near) matter domination and their complete evolution is therefore governed by the physics in the matter era. Finally, ref.~\cite{Follin:2015hya} found a logarithmic dependence of the phase shift on $\ell$ for observable modes.  One may be tempted to interpret this as an effect of the finite matter density. Our goal in this section is to further clarify these effects, by studying the limits $\ell \to \infty$ and $\ell \to 0$, accounting for the contributions from matter. We will study these limits in turn:

\begin{itemize}
\item We first consider modes which entered the horizon during the radiation era. These correspond to small angular scales in the CMB anisotropy spectra.  We begin by writing eq.~(\ref{eq:B})~as
\beq
B(y) =  \underbrace{\int^{y_{\rm eq}}_0 \d y'\, \Phi_+(y')\hskip 1pt\cos y' }_{\displaystyle \equiv B_{\rm rad}} \, \ +\, \ \underbrace{\int_{y_{\rm eq}}^y  \d y'\, \Phi_+(y')\hskip 1pt\cos y'}_{\displaystyle \equiv B_{\rm mat}}  \ ,	\label{eq:BBmatter}
\eeq
where $y_{\rm eq} = c_\gamma k \tau_{\rm eq}$ corresponds to the moment of matter-radiation equality.  Modes that entered the horizon long before $\tau_{\rm eq}$ correspond to $y_{\rm eq} \gg 1$. For these modes, the first term in (\ref{eq:BBmatter}) can be approximated as
\beq
B_{\rm rad} \simeq \int^{\infty}_0  \d y'\, \Phi^{\rm (rad)}_+(y')\hskip 1pt\cos y' \, , 
\eeq
which is precisely the result computed in Section~\ref{sec:analytics}.  The main correction from the matter era then is the second term in (\ref{eq:BBmatter}):
\beq
B_{\rm mat} \simeq \int_{y_{\rm eq}}^y  \d y'\, \Phi_+^{\rm (mat)}(y') \hskip 1pt\cos y' \, . \label{eq:Bmat}
\eeq
To estimate this effect, we simply have to repeat the discussion of \textsection\ref{sec:fs} for the matter era.  The important difference is that $\epsilon_X \equiv \bar \rho_X/\bar \rho$ now is not a constant, but scales as $a^{-1} \propto \tau^{-2}$. Setting $\tau_\in$ in the matter era, we have $\epsilon_X = \epsilon_{X,\in}\hskip 1pt \tau_\in^2/\tau^2$ and eq.~\eqref{eq:Phi-X} becomes
\beq
\Phi_-(y) = - \frac{8k^2}{y^4}  \epsilon_{X,\in} \hskip 1pt y_\in^2\, \sigma_X(y) = - \frac{16}{3}\frac{1}{y^4}  \epsilon_{X,\in} \hskip 1pt y_\in^2\, D_{X,2}(y)  \, . \label{eq:phimmatter}
\eeq
To determine $\Phi_-(y)$ to first order in $\epsilon_X$, we only need the quadrupole moment $D_{X,2}(y)$ to zeroth order. From~(\ref{eq:DX2}), we get
\begin{align}
D_{X,2}^{(0)}(y) \,=\, \ & d_{X,\in} \, j_{2}\!\left[c_\gamma^{-1}(y-y_\in)\right] \nonumber \\
&+ \frac{3}{c_\gamma} \Phi_{+,\in}  \int_{y_\in}^y \d y'\,  \left\{ \frac{2}{5} j_{1}\!\left[c_\gamma^{-1}(y-y')\right] - \frac{3}{5}j_{3}\!\left[c_\gamma^{-1}(y-y')\right]  \right\} \, ,	\label{eq:DX2x}
\end{align}
where we have used that $\Phi_{+}^{(0)} = \mathrm{const.}$ during the matter era. In the limit, $y \gg 1$, this leads~to
\beq
\Phi_-^{(1)}(y) \,\simeq\, 16\epsilon_{X,\in} \, y_\in^2\, \frac{\sin (c_\gamma^{-1} y)}{c_\gamma^{-1}y^5} \Big( \Phi_{+,\in} + \frac{1}{3} d_{X,\in} \Big) \, .
\eeq
Hence, we get $\Phi_- \propto y^{-5}  \to 0$  in the limit $y \to \infty$.  At late times, $\Phi_+$ is therefore no longer sourced by $\Phi_-$ and will be given by the homogeneous solution (with coefficients that may depend on $\epsilon_X$).  Importantly, the value of $B$ will be the same as that predicted in Section~\ref{sec:analytics}.  Hence, at high $k$ (and thus $\ell$), the phase shift of the acoustic oscillations will be equal to the value in a radiation-dominated universe.

\item Next, let us study modes which entered the horizon during the matter era, corresponding to large angular scales in the CMB.  In this case, it is more challenging to cleanly separate the result into a correction to the amplitude of oscillations and a  phase shift.  Our primary goal will be to understand how the result scales with wavenumber $k$ in the limit $k \to 0$.  Fortunately, this scaling is the same for the amplitude correction and the phase shift and is easy to understand analytically.  

Intuitively, we expect the contributions from dark radiation (including neutrinos) to vanish as $k \to 0$.  As we lower $k$, the time of horizon entry increases compared to the time of matter-radiation equality and, therefore, the radiation energy density should be diluted relative to the matter.  Since this radiation only affects observations through its gravitational influence, its role in the evolution of the modes should become negligible.

We can confirm this intuition by returning to~(\ref{eq:phimmatter}) and noticing that $y_\in = c_\gamma k \tau_\in$, where $\tau_\in$ is a fixed time which is independent of $k$, e.g.~we may choose $\tau_\in$ to be the time of matter-radiation equality.  We therefore have $\Phi_-^{(1)} = \epsilon_{X,\in} c_\gamma^2 k^2 \tau_\in^2\, g(y)$, and the correction to $d_\gamma$ at linear order in $\epsilon_X$ will take the form
\beq
d^{(1)}_\gamma (\tau) = \epsilon_{X,\in}\, c_\gamma^2 k^2 \tau_\in^2 \int^{y}_{y_\in} \d y' f(y,y') \, .
\eeq
If the integral converges as $y \to \infty$, it is clear that $d^{(1)}_\gamma \propto k^2 \to 0$.  In fact, if the integral diverges as $\tau \to \infty$, the result will be suppressed by additional powers of $k$, due to the scaling of the upper limit of integration ($y=c_\gamma k \tau$ at fixed $\tau$).  Hence, we conclude that the amplitude and phase corrections from neutrinos (or any dark radiation) will vanish {\it at least} as fast as $k^2$.

\end{itemize}

\noindent
From these asymptotic scaling arguments, we draw the following conclusions:
\begin{itemize}
\item For $k \to \infty$, the phase shift due to free-streaming particles approaches a constant.
\item For $k \to 0$, the phase and amplitude corrections scale at least as $k^2$.
\end{itemize}
 In the flat-sky limit, these results translate approximately to $\ell \simeq  k (\tau_0-\tau_{\rec})$, where $\tau_0$ is the conformal time today.  We therefore expect a constant phase shift at high $\ell$.  Given that matter-radiation equality corresponds to relatively low $\ell$, we do not expect our asymptotic formula for $k \to 0$ to be more than a rough guide. The primary purpose of this discussion was to highlight that a power law is the likely behavior, simply due to the power law decay of the energy density of the extra radiation.  As a result, the phase shift per $\ell$ should be some function that interpolates between a power law and a constant and is unlikely to follow the ansatz of~\cite{Follin:2015hya} in detail (although the logarithmic dependence appears to work well enough on intermediate scales).

%%%%%%%%%%%%%%%%%%%%
\section{Comments on Polarization}
\label{app:pol}
%%%%%%%%%%%%%%%%%%%%
  
The analytic discussions in the main text were phrased in terms of the temperature anisotropy, but as we saw in Section~\ref{sec:analysis}, CMB polarization plays a crucial role in present and future data analysis.  In this appendix, we show that the phase shift of the polarization spectrum is the same as that of the temperature spectrum.

\vskip 4pt
Following~\cite{Zaldarriaga:1995gi}, we write the Boltzmann equation for the amplitude of polarized anisotropies, $\Theta_P$, as
\beq
\dot \Theta_P + i k \mu \Theta_P \,=\, - \dot \kappa  \left[-\Theta_P + \frac{1}{2} \left(1-P_{2}(\mu) \right) \Pi \right] \, ,  \label{eq:A8}
\eeq
where $\Pi \equiv \Theta_{2} + \Theta_{P,0} + \Theta_{P,2}$ and $\dot \kappa  =-n_e \sigma_T a$ is the time derivative of the optical depth~$\kappa$ (to avoid confusion with the conformal time $\tau$). The temperature quadrupole is determined by the photon anisotropic stress, $\Theta_{2} \equiv \frac{1}{2} k^2 \sigma_\gamma $.  Equation~(\ref{eq:A8}) admits a solution as a line-of-sight integral,
\beq
\Theta_P(\tau_0) = \int^{\tau_0}_{\tau_\in} \d\tau \, e^{i k \mu (\tau-\tau_0) - \kappa(\tau)} \left(\frac{3}{4} \dot \kappa(\tau) \hskip 2pt (\mu^2-1)\hskip 2pt \Pi(\tau) \right) \, . \label{eq:ThetaP}
\eeq
The integral in (\ref{eq:ThetaP}) is proportional to the visibility function $-\dot\kappa \hskip 1pt e^{-\kappa}$ and is, therefore, peaked at the surface of last-scattering.  In the limit of instantaneous recombination, $-\dot \kappa e^{-\kappa} \simeq \delta_D(\tau-\tau_\rec)$, we get
\beq
\Theta_P(\tau_0) \simeq e^{i k \mu (\tau_\rec -\tau_0)} \frac{3}{4} \big(1-\mu^2\big) \Pi (\tau_\rec) \, .\label{TP}
\eeq 
Solving for $\Pi$ to leading order in $\dot\kappa \gg 1$, one finds
$\Pi \simeq \tfrac{5}{2} \Theta_2 \simeq -\frac{10}{9} k \, \dot \kappa^{-1} \,\Theta_1$ (using the collision term in the Boltzmann equation for temperature). Applying the continuity equation, $\dot d_\gamma = - 3 k \Theta_1$, and performing a multipole expansion, one finds 
\beq
\Theta_{P,\ell}(\tau_0)\simeq \frac{5}{18} \,\dot d_\gamma(k,\tau_\rec)\, \dot\kappa^{-1}(\tau_\rec)\left(1+ \frac{\partial^2}{\partial (k\tau_0)^2}\right) j_{\ell}(k \tau_0)\, .
\eeq
Two facts should be noticed about this result: 
\begin{itemize}
\item $\Theta_{P,\ell} \propto \dot d_\gamma$.---Since the time derivative will not affect the phase shift from dark radiation, we see that the locations of the acoustic peaks in the polarization spectrum are affected by the fluctuations in the dark radiation in the same way as in the temperature spectrum.
\item $\Theta_{P,\ell} \propto \dot \kappa^{-1} \propto n_e^{-1}$.---This is important because it allows the degeneracy between $H$ and $n_e$ (or $Y_p$) in the damping tail (which scales as $(n_e H)^{-1}$; cf.~\textsection\ref{sec:degeneracies}) to be broken.
\end{itemize}

\clearpage
\phantomsection
\addcontentsline{toc}{section}{References}
\bibliographystyle{utphys}
\bibliography{Refs} 
\end{document}